\documentclass[sigconf]{acmart}

\usepackage{tikz}
\usepackage{subcaption}
\usepackage{multirow}
\usepackage{algorithm}
\usepackage{colortbl}
\usepackage{xspace}
\usepackage{enumitem}

\usetikzlibrary{shapes.geometric, shapes.multipart, arrows, positioning}

\newcounter{insightlabel}
\newcounter{insightnmbr}
\renewcommand{\theinsightlabel}{\textbf{\theinsightnmbr}}
\newenvironment{insight}{
\begin{list}{\textbf{Insight }\theinsightlabel~(\textbf{I}\theinsightlabel):~}{%
\usecounter{insightlabel}\stepcounter{insightnmbr}%
\setlength{\labelwidth}{0pt}\setlength{\labelsep}{0pt}\setlength{\leftmargin}{0in}%
\noindent\rule{\linewidth}{1pt}\item \bf \em}}{\end{list}\noindent\rule{\linewidth}{1pt}}

\definecolor{F1DeepGreen}{HTML}{52b788}
\definecolor{F1MediumGreen}{HTML}{74c69d}
\definecolor{F1LightGreen}{HTML}{95d5b2}
\definecolor{F1LighterGreen}{HTML}{b7e4c7}
\definecolor{F1LightestGreen}{HTML}{d8f3dc}

\definecolor{TimeDeepBlue}{HTML}{00b4d8}
\definecolor{TimeMediumBlue}{HTML}{48cae4}
\definecolor{TimeLightBlue}{HTML}{90e0ef}
\definecolor{TimeLighterBlue}{HTML}{ade8f4}
\definecolor{TimeLightestBlue}{HTML}{caf0f8}

\newcommand{\sysname}{\textsc{FlowWise}\xspace}
\newcommand{\coderepourl}{\url{https://github.com/noise-lab/FlowWise}}

\newcommand{\change}[1]{\textcolor{black}{#1}}
\newcommand{\rchange}[1]{\textcolor{black}{#1}}

\newcommand{\eg}{{\it e.g.}}
\newcommand{\ie}{{\it i.e.}}

\copyrightyear{2026}
\acmYear{2026}
\setcopyright{cc}
\setcctype{by-nc-nd}
\acmConference[SoCC '26]{ACM Symposium on Cloud Computing V.1}{November 18--20, 2026}{Singapore, Singapore}
\acmBooktitle{ACM Symposium on Cloud Computing V.1 (SoCC '26), November 18--20, 2026, Singapore, Singapore}
\acmDOI{10.1145/3815789.3827947}
\acmISBN{979-8-4007-2735-1/2026/11}

\begin{document}
%-------------------------------------------------------------------------------

\title[FlowWise: Stateful Fast-Slow Model Serving]{\texorpdfstring{\sysname: Stateful Fast-Slow Model Serving for Streaming Traffic Intelligence}{FlowWise: Stateful Fast-Slow Model Serving for Streaming Traffic Intelligence}}
% \subtitle{Research Full Paper, \#7, 12 Pages}

\settopmatter{authorsperrow=4, printfolios=true}

\author{Shinan Liu}
\affiliation{%
  \institution{University of Hong Kong}
  \city{Hong Kong SAR}
  \country{China}}
\email{shinan6@hku.hk}

\author{Ted Shaowang}
\affiliation{%
  \institution{University of Chicago}
  \city{Chicago}
  \state{IL}
  \country{USA}}
\email{swjz@uchicago.edu}

\author{Gerry Wan}
\affiliation{%
  \institution{Stanford University}
  \city{Stanford}
  \state{CA}
  \country{USA}}
\email{gwn@stanford.edu}

\author{Jeewon Chae}
\affiliation{%
  \institution{University of Chicago}
  \city{Chicago}
  \state{IL}
  \country{USA}}
\email{jeewon@uchicago.edu}

\author{Jonatas Marques}
\affiliation{%
  \institution{University of Chicago}
  \city{Chicago}
  \state{IL}
  \country{USA}}
\email{jonatas@uchicago.edu}

\author{Sanjay Krishnan}
\affiliation{%
  \institution{University of Chicago}
  \city{Chicago}
  \state{IL}
  \country{USA}}
\email{skr@uchicago.edu}

\author{Nick Feamster}
\affiliation{%
  \institution{University of Chicago}
  \city{Chicago}
  \state{IL}
  \country{USA}}
\email{feamster@uchicago.edu}

\renewcommand{\shortauthors}{Liu et al.}

%-------------------------------------------------------------------------------
\begin{abstract}
%-------------------------------------------------------------------------------
Network traffic analysis in cloud and internet service provider (ISP) networks
increasingly relies on machine learning (ML) models, as encryption and
service consolidation make rule-based classification unreliable. At
\rchange{high flow-arrival rates}, however, new flows can arrive faster than
end-to-end analysis pipelines can process. Unlike typical cloud inference
workloads, packets from the same flow arrive over time and require per-flow
state. \rchange{This streaming, stateful setting limits naive scale-out with
stateless workers.}
This paper presents \sysname, \rchange{a serving system for streaming, stateful
flow intelligence} on commodity CPU/GPU servers that jointly selects (i) how
many packets to collect and (ii) which model(s) to apply for each flow to balance
latency, service rate, and accuracy. Across three representative tasks,
inference time across candidate models differs by 1.8x--141.3x, while packet
waiting time can exceed feature computation and inference by 6--8 orders of
magnitude at the tail.
\sysname exploits these gaps with an uncertainty-driven fast-slow cascade: it
issues early predictions from a one-packet model and escalates only uncertain
flows to slower models that wait for additional packets and use richer features.
A queue-centric runtime maintains per-flow state, performs asynchronous feature
joining, and \rchange{uses brokered, flow-ID-keyed streams to distribute workload across heterogeneous workers.}
\sysname serves 76.3\% of flows within 16 ms, improving median end-to-end
serving latency by 40.5x while increasing service rate and maintaining similar
accuracy to always running the slow model. With thousands of packet-header
features per flow, \rchange{our scale-up benchmark sustains over
48.5k new flows per second on a single 16-core commodity CPU server,} comparable
to reported flow-arrival rates in city-backbone traces.

\smallskip
\noindent\textbf{Code:} \coderepourl.

\end{abstract}




\maketitle
% \fancyhead[RO,RE]{}

\section{Introduction}\label{sec:intro} 

\begin{figure}[t]
  \centering
  \includegraphics[width=0.8\columnwidth]{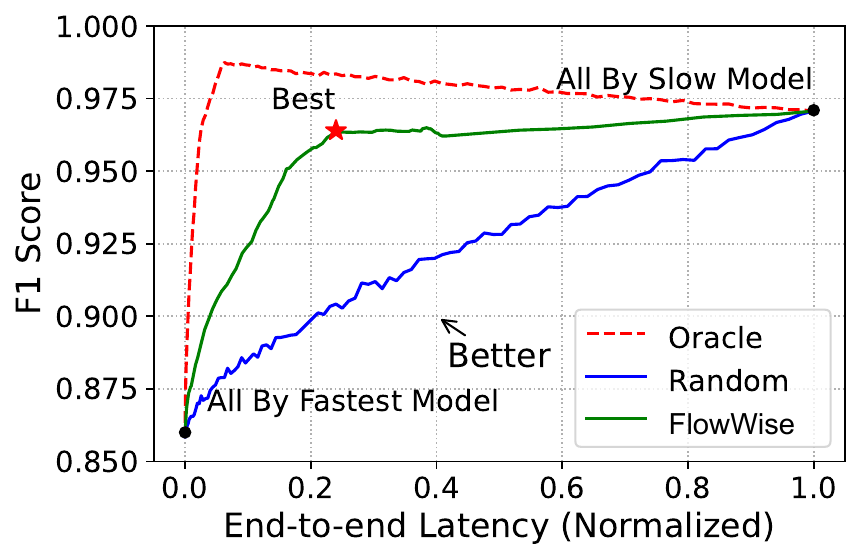}
  \caption{Traffic analysis holds a tradeoff between latency and accuracy. \sysname dynamically assigns predictions to either a slower, more accurate model (waiting for more packets or using a sophisticated design) or a faster, less accurate model (predicting on the first packet). The chart compares its request assignment efficacy against an oracle (using ground-truth classification correctness) and random assignment, based on the \textit{service recognition} dataset.}
  \label{fig:serveflow_teaser}
\end{figure}

Cloud providers and ISPs continuously analyze network traffic to manage
performance~\cite{lotfollahi2020deep, zheng2022mtt, rimmer2017automated,
bronzino2021trafficrefinery,jiang2023acdc}, reliability~\cite{liu2023leaf,
liu2023amir, sharma2023estimating, macmillan2021measuring, piet2023ggfast}, 
and security~\cite{holland2021new, jiang2023generative, jiang2023netdiffusion}. 
This analysis increasingly must operate
on encrypted traffic and at multi-tenant scale, where payload inspection and
simple heuristics are unreliable. As a result, operators
have turned to data-driven models to classify and characterize traffic
in real time, by looking into their raw-packet headers~\cite{holland2021new} or flow-level 
statistics~\cite{jiang2023acdc,rimmer2017automated,swamy2023homunculus}.

Many of these models support
network traffic analysis tasks, such as service
recognition~\cite{bernaille2006earlyappid,shapira2021flowpic,
rezaei2019deepappid,jiang2023acdc}, quality of experience (QoE) measurement on
encrypted traffic~\cite{mangla2018emimic,
bronzino2019inferring,sharma2023estimating}, intrusion
detection~\cite{khraisat2019survey,ahmad2021network,liu2019machine} and device
identification~\cite{yang2021efficient,sivanathan2019iotclass}. 
The output of these tasks is often consumed by cloud and ISP control loops (e.g.,
prioritization, rate limiting, tenant isolation, and incident response),
\rchange{which require timely actions while flows are still active}.
For flow prioritization and QoE control, service recognition and QoE
measurement must finish within tens of milliseconds, often under 50
ms~\cite{staessens2011software}. \rchange{In VCA traffic, WebRTC measurements
show that 350--380 ms tail delays can cause about 12\% audio concealment and
seconds of video freeze~\cite{yi2025domino}. Security tasks impose the same
deadline: malicious payloads or control signals must be blocked before
completion~\cite{ahmad2021network,mittal2023deep}, and a one-second DDoS
detection delay can admit 47.5 GB of attack traffic at 380
Gbps~\cite{liu2021jaqen}.} Although ML-based approaches
are typically more accurate and robust than heuristics, deploying
them in high-bandwidth provider networks is challenging: \rchange{the value of a
traffic label decays with delay, because these actions are only useful when they
can still shape the ongoing flow or nearby traffic.} The analysis service
must keep up with the incoming packet stream while producing decisions that are
timely enough to affect active flows, even at a traffic rate of tens or hundreds
of Gbps in typical cloud settings.

The community has explored two main approaches to run ML-based traffic analysis
at \rchange{high flow-arrival rates}. The first accelerates inference by moving feature extraction and
prediction into \textbf{specialized hardware} such as SmartNICs or smart switches (\eg,
N3IC~\cite{siracusano2022n3ic}, Homunculus~\cite{swamy2023homunculus},
LEO~\cite{jafri2024leo}, NetBeacon~\cite{zhou2023efficient}, and
BoS~\cite{yan2024brain}). Although these systems can deliver high throughput,
they are often a poor fit for \rchange{provider deployments}, where operators prefer
solutions that are easy to deploy and do not require changes to the data plane.
Specialized hardware restricts supported operations (\eg, floating-point
arithmetic, dot products, and divisions), limits the model families that can be
deployed (\eg, binary neural networks~\cite{swamy2022taurus}, decision trees,
random forests~\cite{bussegrawitz2019pforest}, and small
RNNs~\cite{yan2024brain}), and typically requires domain-specific toolchains,
which together increase engineering cost and reduce opportunities for flexible
scale-out. The second approach keeps inference on \textbf{commodity CPU/GPU servers} and
instead focuses on reducing per-flow cost through efficient featurization and
model selection; for example, CATO~\cite{wan2025cato} uses Bayesian optimization to search for
cost-effective feature sets, AC-DC~\cite{jiang2023acdc} relies on profiling to
select models, and LEXNet~\cite{fauvel2023lightweight} reduces computation with
a lightweight CNN over packet size and direction sequences. These approaches 
can be effective, but \rchange{many rely on compact feature representations},
which limit the effectiveness and generalizability \rchange{for deployments that
need richer packet-header information.}

A key insight that both lines of work neglected is that, simply reducing the latency of \textbf{featurization} and \textbf{model inference}, 
may not sufficiently improve the performance of the end-to-end system. 
The initial \textbf{flow collection step can contribute orders of magnitude more 
latency} because the model must wait for packets to arrive.
To understand these deployment bottlenecks, consider a typical network 
classification workflow (Fig.~\ref{fig:netml_pipe}). First, a packet-capturing framework collects and 
segments active network flows (flow collection). Next, these flows are converted into numerical features (featurization). These features are 
inputs to the model which returns a classification result (inference). 
\textit{In many ML problems}, the primary bottleneck is the inference step.
There are many techniques to improve model inference latency.
These include model pruning and quantization~\cite{liu2018rethinking,fang2023depgraph,
fan2020training}, knowledge distillation~\cite{wang2021knowledge,
gou2021knowledge}, hardware acceleration~\cite{ghimire2022survey}, 
data compression~\cite{yang2020improving,hu2020fast}, and model 
specialization~\cite{li2021dynamic,hu2021lora}. 
Accordingly, there are several approaches applied to \textit{networking problems} 
including: efficient modeling~\cite{piet2023ggfast,xu2023fasttraffic,fauvel2023lightweight,
koksal2022markov, qiu2022traffic, tong2014high, devprasad2022context, liu2019adaptive}, 
feature selection~\cite{bronzino2021trafficrefinery, jiang2023acdc}, 
and specialized hardware (like SmartNICs)~\cite{siracusano2022n3ic, 
swamy2023homunculus, jafri2024leo, zhou2023efficient,yan2024brain}.

% Let's further analyze the data collection costs.
For most practical use cases, the classifier needs to be able to issue a traffic 
classification while a flow is active so the system can act on this result (e.g., 
change flow priority or reject malicious flows). However, the model must wait for sufficient information 
from a flow to arrive before doing any further processing. 
Waiting longer gives the model more context on a flow, and thus, a more 
accurate final result~\cite{bernaille2006earlyappid, piet2023ggfast}.
Conventional flow-level classification methods typically wait between 4 and 100 
packets~\cite{holland2021new,piet2023ggfast,jiang2023acdc}, while others demand 
the entire flow~\cite{shapira2021flowpic,bronzino2019inferring,sharma2023estimating}. For example, 
a recent high-performance method GGFAST~\cite{piet2023ggfast} 
needs to wait for 50 packets to craft informative sequence-of-lengths features. 
\change{LEXNet~\cite{fauvel2023lightweight} collects the sizes and directions of 20 packets as features.}
Waiting for packets from a particular flow to arrive is a form of model latency. 
It requires a dedicated system thread to query the network capture stream 
which reduces the overall throughput of a classification system.
% \rchange{In this setting, latency and throughput are tightly coupled: waiting for
% more packets both delays the decision and keeps per-flow state alive longer,
% which lowers service rate under load.}

\begin{figure}
  \centering
  \begin{tikzpicture}[
  node distance=1cm and 0.5cm, 
  auto,
  block/.style={
    rectangle, 
    draw=black, 
    fill=white, 
    align=center,
    rounded corners,
    text width=0.2\columnwidth, 
    minimum height=2em
  },
  smallblock/.style={
    rectangle, 
    draw=black, 
    fill=white, 
    align=center,
    rounded corners,
    text width=0.205\columnwidth, 
    minimum height=2em,
    font=\small
  }, 
  bigblock/.style={
    rectangle, 
    draw=black, 
    fill=white, 
    align=center,
    rounded corners,
    text width=0.5\columnwidth, 
    minimum height=5em
  },
  line/.style={
    draw, 
    -latex'
  }
  ]
  % \todo{full packet classification can work better, but it is super hard to make it run at line rate}
  % Nodes
  \node [anchor=west] (traffic) at (0,0) {Traffic};
  \node [bigblock, right=of traffic] (bigfeature) {};
  \node [smallblock, anchor=south west, yshift=0.2cm, xshift=0.2cm] at (bigfeature.south west) (compute) {Flow \\ Collection};
  \node [smallblock, anchor=south east, yshift=0.2cm, xshift=-0.2cm] at (bigfeature.south east) (flow) {Feature Computation};
  \node [below=0.1cm of bigfeature.north] {Feature Extraction};
  \node [block, right=of bigfeature] (model) {Model Inference};  
  
  % Paths
  \path [line] (traffic) -- (bigfeature);
  \path [line] (bigfeature) -- (model);
  
  \end{tikzpicture}
  \caption{Model serving for network traffic analysis.}
  \label{fig:netml_pipe}
  \vspace{-1em}
\end{figure}

Although this gap in flow collection time might seem like a fundamental technical 
bottleneck, this paper contributes a key insight into many network traffic 
analysis problems: some flows are easy to classify, and some are harder.
For example, even for a single packet (as modeled by many packet-level traffic 
analytical frameworks such as Homunculus~\cite{swamy2023homunculus}, LEO~\cite{jafri2024leo}, etc.), 
a typical web browsing request may display standard TCP options like MSS, and SACK, while 
specialized applications or services like VPNs or database query protocols often 
exhibit unique or rare TCP options. In these
cases, it is possible to accurately classify some traffic with a single packet.
To exploit this skew, we develop \rchange{a serving system for streaming,
stateful flow intelligence that jointly adapts packet depth and model choice}.
Instead of using a single conventional ``slow'' (requires long flow context but is more accurate) model, 
we place additional ``fast'' (requires less flow context) models in front of it. 
This architecture can balance accuracy 
with waiting time (i.e., return a less accurate answer earlier).
A selection algorithm assigns uncertain predictions from the fast model 
for further processing on the slow model. 
\rchange{The system contribution is not a generic layered cascade. It is a
stateful serving design for streaming network flows across asynchronous packet
arrivals.}
This fast-slow architecture is 
especially beneficial when the two models involved have high disparities 
in their performance and operational cost.
Figure~\ref{fig:serveflow_teaser} illustrates the basic results, showing that \sysname presents 
the user with a latency-accuracy tradeoff rather than a single design point. 

\begin{table}[t]
\centering
\footnotesize
\begin{tabular}{llccc}
\toprule
\textbf{Application}                                                             & \textbf{Model} & \textbf{F1 (1pkt)} & \textbf{F1(5pkts)} & \textbf{F1(10pkts)} \\ \hline
\multirow{3}{*}{\begin{tabular}[c]{@{}l@{}}Service\\ Recognition\end{tabular}}   & DT             & $0.845$            & $0.907$            & $0.909$             \\
                                                                                 & LGBM           & $0.921$            & $0.967$            & $0.973$             \\
                                                                                 & CNN            & $0.887$            & $0.929$            & $0.930$             \\ \hline
\multirow{3}{*}{\begin{tabular}[c]{@{}l@{}}Device\\ Identification\end{tabular}} & DT             & $0.812$            & $0.824$            & $0.827$             \\
                                                                                 & LGBM           & $0.913$            & $0.918$            & $0.919$             \\
                                                                                 & CNN            & $0.881$            & $0.910$            & $0.911$             \\ \hline
\multirow{3}{*}{QoE Inference}                                                   & DT             & $0.715$            & $0.740$            & $0.745$             \\
                                                                                 & LGBM           & $0.723$            & $0.769$            & $0.773$             \\
                                                                                 & CNN            & $0.710$            & $0.752$            & $0.764$             \\
\bottomrule
\end{tabular}
\caption{Empirically, adding more packets increases the (weighted) F1 score. DT: Decision Tree, LGBM: LightGBM, CNN: Convolutional Neural Networks.}
\label{tab:f1_scores}
\vspace{-2em}
\end{table}

\sysname enables \rchange{\textbf{high-rate ML-based encrypted traffic analysis
on commodity servers}}, without specialized hardware. It
supports expressive packet-header representations and lets operators train models in Python-based ML
libraries and deploy them through a common serving interface, enabling short development cycles for operators. \sysname makes four major contributions.
\begin{itemize}
  \item We characterize end-to-end latency in online traffic analysis pipelines and show that packet waiting time can exceed model inference time by several orders of magnitude, even when using high-dimensional packet-header features.
  \item We design a stateful fast-slow serving architecture that jointly selects packet depth and model placement, and we develop uncertainty-based flow assignment algorithms that escalate only uncertain flows to slower models.
  \item We implement \sysname as a queue-centric serving runtime that maintains per-flow state and supports asynchronous feature joining, \rchange{using brokered, flow-ID-keyed streams to support scaling across computing resources.}
  \item We open source an easy-to-use development and deployment ML-based traffic analysis library for cloud vendors 
  and ISPs: \sysname can serve models trained in Python-based ML libraries without 
  requiring specialized hardware or domain-specific languages.
\end{itemize}

Across three traffic-analysis tasks, \sysname achieves 40.5$\times$ lower
median end-to-end latency with comparable accuracy and sustains over 48.5k new flows/s on a 16-core commodity server.

\begin{figure*}[!t]
  \centering
  \begin{minipage}{0.52\textwidth}
    \centering
    \subfloat[Service Rec. \label{fig:serv_rec_cdf}]{%
      \includegraphics[width=0.33\textwidth]{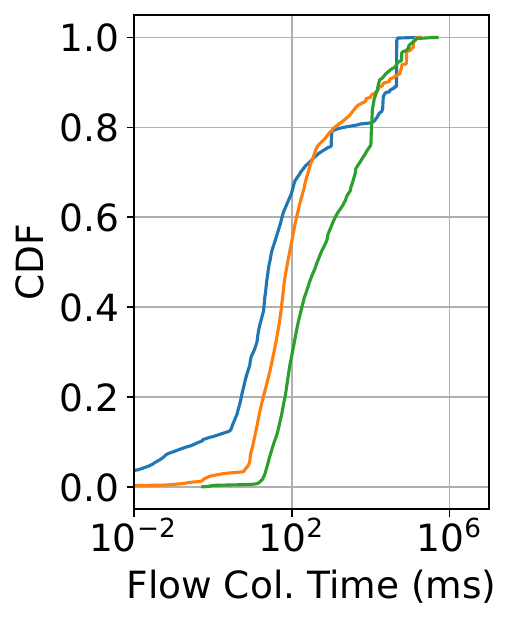}%
    }\hfill
    \subfloat[Device Ident. \label{fig:dev_id_time_cdf}]{%
      \includegraphics[width=0.33\textwidth]{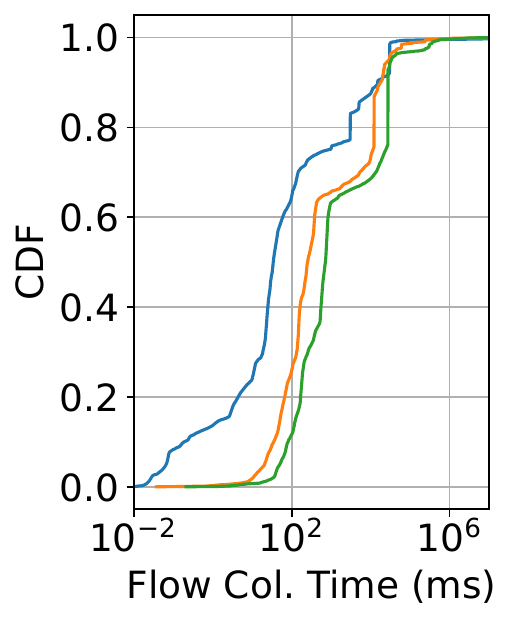}%
    }\hfill
    \subfloat[QoE Inference. \label{fig:vca_qoe_time_cdf}]{%
      \includegraphics[width=0.33\textwidth]{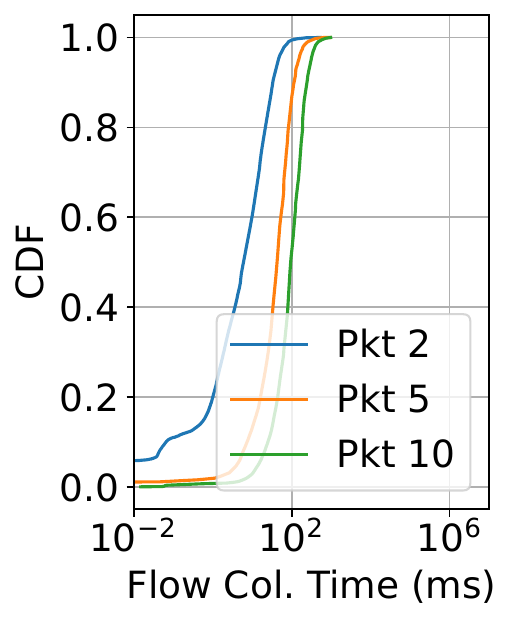}%
    }
    \caption{Flow collection time (since the arrival of the first packet, in ms) across 
    applications. Note that some flows are short-lived, which do not have five (or more) 
    packets. So at the tail, flow collection time would be smaller compared 
    to collecting two packets. }
    \label{fig:time_cdf}
  \end{minipage}%
  \hfill
  \begin{minipage}{0.465\textwidth}
    \centering
    \small
    \begin{tabular*}{\textwidth}{@{\extracolsep{\fill}}c l c c c c@{}}
    \toprule
    \textbf{Stage} & \textbf{Application} & \textbf{Model} & \textbf{1 Pkt} & \textbf{5 Pkts} & \textbf{10 Pkts} \\
    \hline
    \multicolumn{2}{c}{Feature Computation} & - & 0.012 & 0.058 & 0.139 \\
    \hline
     & \multirow{3}{*}{Service Rec.} & DT & 0.052 & 0.142 & 0.251 \\
    & & LGBM & 0.139 & 0.237 & 0.358 \\
    & & CNN & 0.481 & 3.815 & 7.345 \\
    \cline{2-6}
    Model & \multirow{3}{*}{Device Id.} & DT & 0.053 & 0.145 & 0.275 \\
    Inference & & LGBM & 0.210 & 0.322 & 0.460 \\
     & & CNN & 0.491 & 3.277 & 6.446 \\
    \cline{2-6}
    & \multirow{3}{*}{VCA QoE} & DT & 0.053 & 0.095 & 0.164 \\
    & & LGBM & 0.162 & 0.218 & 0.298 \\
    & & CNN & 0.385 & 3.290 & 6.543 \\
    \bottomrule
    \end{tabular*}
    \captionof{table}{Median feature compute time, and median inference time (both in ms) for different models. 
    Both numbers for the 1st packet features are in the range of 
    $10^{-2}$ to $10^{-1}$ ms. DT: Decision Tree, LGBM: LightGBM, CNN: Convolutional Neural Networks.}
    \label{tab:time_bottleneck}
\end{minipage}
\end{figure*}

% \begin{table}[!t]
%   \centering
%   \small
%   \begin{tabular}{llccc}
%   \toprule
%   \textbf{Application}     & \textbf{Model}    & \textbf{F1 (1pkt)}  & \textbf{F1(5pkts)} & \textbf{F1(10pkts)} \\ 
%   \midrule
%                           & DT            & $0.845$ & $0.907$ & $0.909$ \\
%   Service                & LGBM          & $0.921$ & $0.967$ & $0.973$ \\
%   Recognition              & CNN           & $0.887$ & $0.929$ & $0.930$ \\
%   \midrule 
%       & DT            & $0.812$ & $0.824$ & $0.827$ \\
%   Device                & LGBM          & $0.913$ & $0.918$ & $0.919$ \\
%   Identification              & CNN           & $0.881$ & $0.910$ & $0.911$ \\
%   \midrule             
%                     & DT            & $0.715$ & $0.740$ & $0.745$ \\
%   QoE Inference          & LGBM          & $0.723$ & $0.769$ & $0.773$ \\
%                            & CNN           & $0.710$ & $0.752$ & $0.764$ \\
%   \bottomrule
%   \end{tabular}
%   \caption{Empirically, adding more packets increases the classification (weighted) F1 score. DT: Decision Tree, LGBM: LightGBM, CNN: Convolutional Neural Networks.}
%   \label{tab:f1_scores}
% \end{table}

\section{Background and Motivation}\label{sec:motivation}

Next, we describe the functionality of existing ML model serving systems (and approaches) and discuss why they are insufficient for modern network traffic analysis.

\subsection{General-purpose Model Serving Systems}

Serving trained models is an important part of any ML task.
Model serving systems, such as TensorFlow Serving~\cite{olston_tensorflow-serving_2017}
and TorchServe~\cite{paszke2019pytorch}, provide interfaces for serving previously trained models.
These systems are designed for scaling out in the cloud and prioritize ``parallelism through asynchrony'', assuming each new data input is independent and represents a complete inference request that can be served by any in a pool of threads.
Indeed, many model-serving frameworks (\eg, Clipper~\cite{crankshaw2017clipper}, 
InferLine~\cite{crankshaw2020inferline}) leverage large numbers
of parallel workers to improve overall system throughput. 
When data inputs are independent of each other, such 
a stateless-parallelism approach is straightforward.
However, the streaming nature of network traffic means that each individual packet (data input) may need to be coupled with other packets (from the same flow) to enable proper context for inference.
Thus, in the network setting, there is additional state management needed to correctly situate packets within flows in order to assemble complete inference requests.
This poses an additional challenge on top of the fact that network traffic rate and volume can often exceed the maximum capacity of workers.
In this context, it is essential to develop systems capable of intelligently prioritizing network flows to gracefully degrade inference performance when resources are saturated.
For these reasons, we develop \sysname, a new model serving framework focused on network traffic analysis problems.

\subsection{Machine Learning for Traffic Analysis}
Many machine learning approaches for network traffic analysis are designed for
the offline setting, where models are applied to previously collected flows.
Such approaches require observing either a few packets (usually between 4 and
100) or complete flows for effective classification~\cite{holland2021new,
piet2023ggfast,jiang2023acdc, shapira2021flowpic,bronzino2019inferring,
sharma2023estimating}. \change{In the online setting, even though systems like
Homunculus~\cite{swamy2023homunculus}, Taurus~\cite{swamy2022taurus},
LEO~\cite{jafri2024leo}, and BoS~\cite{yan2024brain} support inferences at line
rate, \textit{they mainly optimize for the feature computation time or the model inference 
time}. They either conduct per-packet inference or wait to buffer new packets
from a flow. The former method leads to repetitive computation for flow-level
analysis, while the latter adds prohibitive latency to an ML-based networking
component. In addition, these systems require specialized hardware like
SmartNICs, or substantial re-architecture of such hardware platforms to achieve
line-rate inferences. As a result, these platforms often do not support existing
code libraries for GPU/CPU, limiting the model architectures that can be served. Furthermore, the 
features contemplated in these systems are often limited to simple flow-based 
statistic features, which cannot capture the complexity of modern network 
traffic as much as raw-packet features~\cite{holland2021new,mittal2023deep}.} 

\change{To further explore opportunities for enhancing ML-based network traffic
analysis, next, we reassess the bottlenecks in three different tasks: service
recognition, device identification, and (video conferencing application, \ie, VCA) 
QoE inference (detailed in Appendix~\ref{app:implementation-details}). Our investigation yields two key insights.}

\begin{insight} 
  \change{Despite improving model performance, waiting for additional context
  leads flow collection time to be several orders of magnitude longer
  than inference time.}
\end{insight}
There are two key tradeoffs at play here. On the data side, longer context
lengths (i.e., observing more data for a given flow) often enable training
more accurate models but also suffer from higher end-to-end latency. This
latency comes from both packet waiting time and additional feature processing
time. On the model side, more expressive models can be more accurate but come
with higher inference latencies and may require more computing resources (\eg,
higher RAM usage or specialized hardware). Table~\ref{tab:f1_scores}
and~\ref{tab:time_bottleneck} put these tradeoffs into perspective. They show
the F1 score and latency as a function of context size (i.e., number of packets per flow)
for several widely-used models~\cite{lotfollahi2020deep, tong2014high,
devprasad2022context,holland2021new,liu2023amir} and three different inference
tasks. \textit{The accuracy increases as we increase the number of packets per flow, 
as do feature computation and model inference times}.

For online flow classification, in addition to the aforementioned delays, 
we would also have to wait for packets to arrive before prediction of the flow 
can take place. \textit{We find that the end-to-end latency is generally dominated by flow 
collection time.} 
Figure \ref{fig:time_cdf} shows the CDF of waiting times for various packet
depths. Waiting for additional packets increases the time: the median waiting
time for the 2nd packet ranges from several to tens of milliseconds, while for
the 10th packet it ranges from hundreds to thousands of milliseconds, depending
on the application.
\change{Further, contrasting Fig.~\ref{fig:time_cdf} with Table~\ref{tab:time_bottleneck}, 
we can see that the model inference latency is orders of magnitude smaller than 
the packet waiting time for three different types of models (detailed in Sec.~\ref{subsec:setting}).
For example, to wait for the second packet, the median flow collection time 
is around tens of ms while the maximum time is between $10^3$ and $10^6$ ms.
In contrast, the feature computation time and model inference time are both 
stable across flows, as they are often derived from dense linear 
algebra operations with minimal control flow, and in the range of 
$10^{-2}$ to $10^{-1}$ ms for a single packet context.
This indicates that even before the second packet arrives, a system can 
have more than enough time to derive inferences for each flow.
}

\rchange{Note that this effect is not caused by constrained throughput. For each flow, our
measurements compare the time to wait for later packets against feature
computation and inference time. The gap comes
from per-flow arrival dynamics, not from saturating the system. We observe this
across datasets with different traffic patterns, including streaming traffic and
VCA, where flows are longer-lived and packets arrive relatively quickly.}

\begin{insight}
  \change{For the same traffic analysis task, inference time across models with different architectures and/or context lengths can differ by orders of magnitude.}
\end{insight}

\change{
According to Table~\ref{tab:time_bottleneck}, inference times can vary significantly across models for the same task.
In the table, these variations range from 1.8x to 141.3x depending on factors like architecture and context length.
For example, for service recognition, the inference time difference between using a Decision Tree on the first packet and a CNN on the first 10 packets is as high as 141.3x. 
For device identification, different modeling on the first packet can have up to 9.3x difference (Decision Tree vs. CNN) in inference time.
Furthermore, we also compare the model performance in F1 score with 
the latency of model inference on the first packet of each flow (Table~\ref{tab:1pkt_f1vs.inf}). 
From Table~\ref{tab:f1_scores},~\ref{tab:time_bottleneck} and~\ref{tab:1pkt_f1vs.inf}, interestingly, 
the model with the highest F1 score (LightGBM) is not the slowest, while the 
fastest model (Decision Tree) delivers an acceptable F1 score. This highlights 
a trade-off between latency and accuracy, suggesting a valuable search space for optimizing model selection.
}

\begin{table}[!t]
  \centering
  \footnotesize
  % \footnotesize
  \begin{tabular}{lcccccc}
    \toprule
    & \multicolumn{2}{c}{\textbf{Serv. Rec.}} & \multicolumn{2}{c}{\textbf{Dev. Id.}} & \multicolumn{2}{c}{\textbf{QoE Inf.}} \\
    \textbf{Model} & \textbf{F1} & \textbf{Inf. Time} & \textbf{F1} & \textbf{Inf. Time} & \textbf{F1} & \textbf{Inf. Time} \\
    \midrule
    DT & \cellcolor{F1LighterGreen}0.845 & \cellcolor{TimeDeepBlue}0.052 & \cellcolor{F1LighterGreen}0.812 & \cellcolor{TimeDeepBlue}0.053 & \cellcolor{F1MediumGreen}0.715 & \cellcolor{TimeDeepBlue}0.053 \\
    RF & \cellcolor{F1LightGreen}0.859 & \cellcolor{TimeMediumBlue}0.107 & \cellcolor{F1LightGreen}0.874 & \cellcolor{TimeMediumBlue}0.106 & \cellcolor{F1LighterGreen}0.709 & \cellcolor{TimeMediumBlue}0.098 \\
    LGBM & \cellcolor{F1DeepGreen}0.921 & \cellcolor{TimeLighterBlue}0.139 & \cellcolor{F1DeepGreen}0.913 & \cellcolor{TimeLighterBlue}0.210 & \cellcolor{F1DeepGreen}0.723 & \cellcolor{TimeLighterBlue}0.162 \\
    XGBoost & \cellcolor{F1LightestGreen}0.672 & \cellcolor{TimeLightBlue}0.117 & \cellcolor{F1LightestGreen}0.507 & \cellcolor{TimeLightBlue}0.152 & \cellcolor{F1LightestGreen}0.646 & \cellcolor{TimeLightBlue}0.127 \\
    CNN & \cellcolor{F1MediumGreen}0.887 & \cellcolor{TimeLightestBlue}0.481 & \cellcolor{F1MediumGreen}0.881 & \cellcolor{TimeLightestBlue}0.491 & \cellcolor{F1LightGreen}0.710 & \cellcolor{TimeLightestBlue}0.385 \\
    \bottomrule
  \end{tabular}
  \caption{Different model performance for the 1st packet. Deeper green means higher F1 score, and deeper blue means lower model inference latency (in ms).}
  \label{tab:1pkt_f1vs.inf}
  \vspace{-2em}
\end{table}

The primary goal of \sysname is to enhance performance by reducing inference latency while preserving the accuracy of its predictions.
Thus, we develop unique strategies to (1) apply a novel fast-slow serving architecture on both partial data (\change{Insight 1,} the first packet of a network flow) and how to use this partial data (\change{Insight 2,} by carefully choosing models to balance system cost and model performance); (2) design a mechanism for assigning flow classification requests to different models; and (3) fully leverage diverse resources available in heterogeneous hardware environments.

\section{\rchange{Stateful} Fast-Slow Modeling}\label{sec:system}
In this section, we explore the rationale and requirements for \rchange{a serving system
for streaming, stateful network traffic}, introducing \sysname which optimizes it 
through a Fast-Slow modeling. The key to 
this architecture is an algorithm developed to effectively differentiate between correct and incorrect predictions. We detail two designs in this section.

\subsection{Fast-Slow Model Architecture}\label{sec:fast-slow-arch}
In any network classification problem that operates on flows, the amount of context provided to the model is a key parameter, namely, how much of the flow can the model see before making a decision.
In prior work, this has been treated as a fixed hyper-parameter, but we find that varying this information to influence the responsiveness of the system is valuable.
For instance, service 
recognition models can learn the service categorization of a flow using only the TCP options in the first packet~\cite{holland2021new}.
Some categorizations may need more information to distinguish. They need 
inter-arrival time, sequences of packet sizes, or other flow-level features
to provide more context. For example, QoE measurement on encrypted traffic needs 
more packets from the same session to understand rebuffering ratio~\cite{mangla2018emimic} or frame
rates~\cite{sharma2023estimating}.

% \noindent \textbf{Why might a fast-slow architecture be beneficial?}
% Taking advantage of the inherently discrete arrival of these packets, we have designed a
% fast-slow architecture for traffic analysis. This architecture enables inference based on partial flow
% information (i.e., limited context), addressing the varying demands for speed and accuracy. While slow
% models typically yield more accurate predictions, their reliance on the
% accumulation of a larger number of packets or on extensive computation makes
% them time-intensive. Conversely, a fast model offers rapid, albeit less accurate,
% predictions. Acting as a ``filter'', the fast model initially processes all
% traffic, forwarding only flows that necessitate higher accuracy to the
% slower models. For those satisfactory predictions made by the 
% fast model, the latency to decision is reduced by a large margin because it avoids waiting 
% for more packets to come and more inference time.

\noindent \textbf{When would a fast-slow architecture be beneficial?} Fast-slow modeling
becomes particularly beneficial in scenarios where there is a significant
disparity in operational costs between the fast and slow models. This encompasses factors such as data collection delays, featurization costs, state information tracking, and computational or memory demands for inference. Additionally, it also requires a sufficient gap
in model performance: in some instances, even a fast model can meet the accuracy
requirements for certain classes, making a slow model unnecessary.
For many ML tasks, the complexity of a model often highly correlates with its
operational cost, and it can also be a tentative indicator of the model's
performance. Thus, it forms an interesting search space for efficiency 
improvements. 

\noindent \textbf{How might the fast-slow model architecture generalize?} As
discussed above: for the same task, if two models share
disparities across their costs and performances, a fast-slow architecture could
apply. For example, in network traffic analysis, two such opportunities exist:
Firstly, packet waiting time often exceeds model inference time by several orders of 
magnitude, creating a natural disparity in latency. Secondly, the time taken
for inference across identical inputs (whether partial or full) can also vary greatly,
sometimes by 10x or more, as presented in Table~\ref{tab:time_bottleneck}. This understanding opens the door to potentially even
more nuanced applications of the architecture. For instance, a multi-tiered
fast-slow approach could be implemented, where adjustments are made between each
packet arrival (considering that packet inter-arrival times often surpass model
inference times) or for different models at each packet number threshold. The
challenge is ensuring that the time saved outweighs added management overhead.
\change{We believe this architecture can be extended beyond 
network analysis, particularly in cases where two models exhibit significant 
operational costs and performance discrepancies.}

\begin{figure}[!t]
    \centering
    \includegraphics[width=\columnwidth]{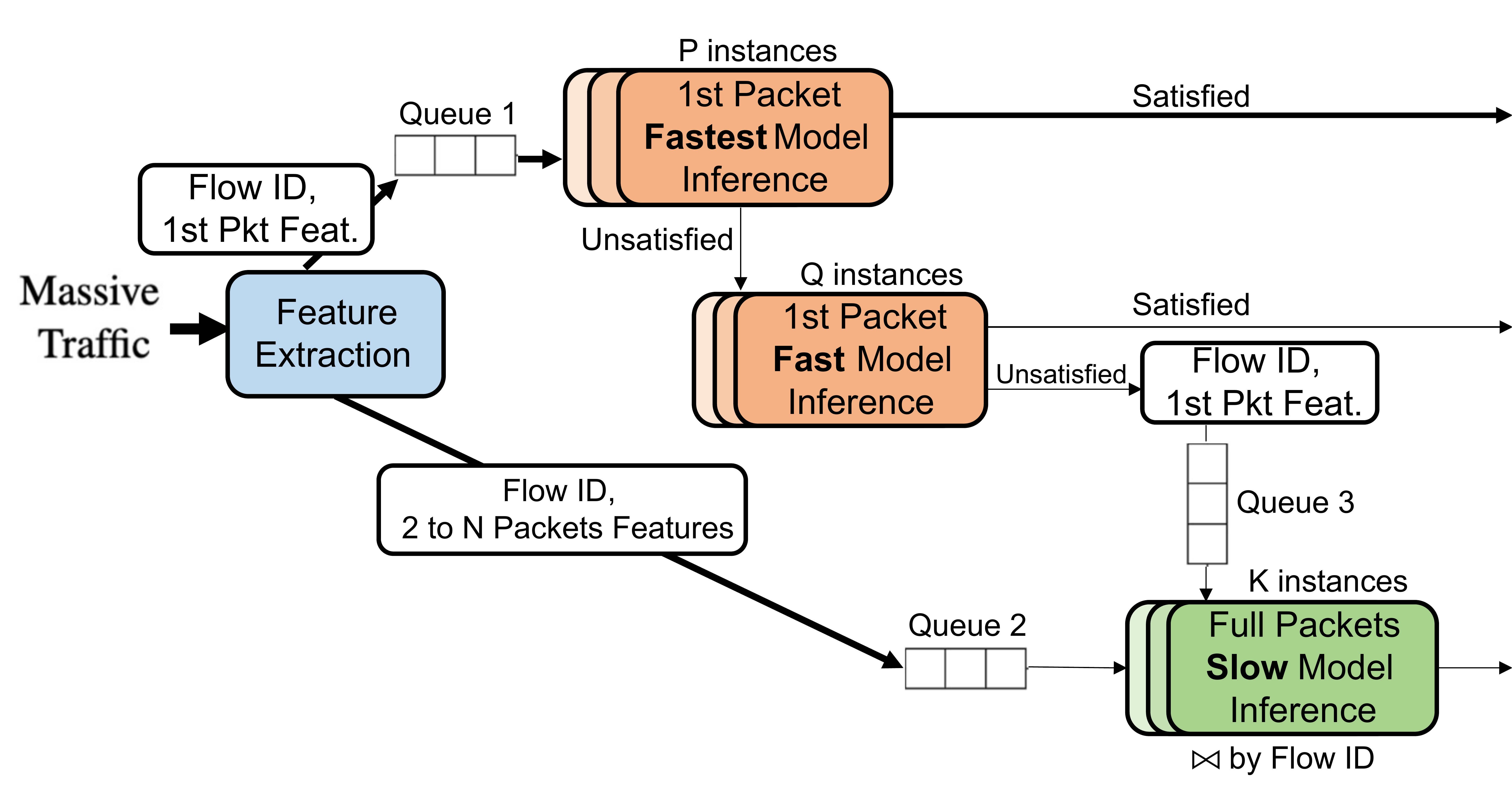}
    \caption{\sysname, which supports the 
    fast-slow model serving architecture, is tailored to the discrete arrival of network flow packets. The arrow width represents 
    the proportion of flows processed at each stage. }
    \label{fig:serveflow}
    \vspace{-0.6em}
  \end{figure}

We propose \sysname, where we capitalize on these two optimization avenues: the
number of early-stage packets \change{(Insight 1)} and the choice of model \change{(Insight 2)}. As depicted in
Figure~\ref{fig:serveflow}, the ``fastest'' and ``fast'' model both use features only from the first packet, and the ``slow'' model uses more packets as input.
Each flow is uniquely identified by its 5-tuple (source IP and port, destination IP and port, protocol), referred to as a flow identifier (flow ID).
Upon a flow's entry into the network, the first
packet is converted into an nPrint feature representation~\cite{holland2021new}, which 
effectively embeds information of all the raw header field bits. 
It is then sent to Queue~1 for immediate processing by the fastest inference model.
This stage quickly filters through the traffic, identifying flows that are
potentially misclassified. Should the fastest model's
predictions be deemed insufficient, the same initial packet features are
subsequently analyzed by the fast model.
% This assesses whether a full packet inspection is necessary, offering a more refined
% evaluation without significantly compromising on speed.
The fastest model,
typically several times quicker than the fast model, significantly enhances
throughput and reduces latency, while the fast model helps assess whether more context is necessary.
\change{If the prediction from the fast model is still unsatisfactory, the data
is forwarded to the slow model via Queue~3. Meanwhile, features from subsequent
packets along with their flow IDs are accumulated in Queue~2. The slow model
pulls data from both Queue~2 and Queue~3 in real time, combining these features
based on flow ID to perform inference. Since the fastest and fast models provide
satisfactory predictions for the majority of flows, only a small portion is
assigned to Queue 3.
Unused data accumulated in Queue~2 and Queue~3 without a flow ID match in the other queue is discarded after a timeout.}
\rchange{\sysname can therefore be viewed as a slow-path baseline with a
fast-path and early-exit accelerator. If unusual traffic pushes more flows to
the slow path, \sysname degrades toward the conventional packet-waiting regime
rather than failing. Threshold selection and timeout-based cleanup manage this
slow-path pressure.}
%  \todo{we also talk about the buffer management in section 4.1. However, we haven't talked about buffer size and how our data retention policy impacts accuracy.}
This asynchronously paced approach effectively balances the trade-off
between speed and accuracy, ensuring quick response times for initial traffic
assessment and detailed analysis for assigned classification requests. 
Note that not all \sysname applications necessitate a fastest-fast-slow 
architecture, as the cost and performance gaps might not be large 
enough on one of the dimensions.

\rchange{Aside from fixed-length ML models, a single streaming or variable-length model 
can also be an alternative model
family, not a replacement for the system problem that \sysname addresses.
Streaming methods such as Homunculus~\cite{swamy2023homunculus} discard
per-packet structural detail, while raw packet features such as nPrint preserve
header-level information that improves downstream task performance~\cite{holland2021new}. RNNs and
Transformers over packet sequences could also be used inside \sysname: they
would still need per-flow state management across asynchronous arrivals and
early exit for confident flows.}

% \begin{figure}[!h]
%     \centering
%     \subfloat[Service Recognition \label{fig:serv_rec_pareto}]{%
%     \includegraphics[width=0.322\textwidth]{figures/serv_rec_pareto.pdf}%
%     }\hfill
%     \subfloat[Device Identification \label{fig:dev_id_time_pareto}]{%
%     \includegraphics[width=0.322\textwidth]{figures/device_identification_pareto.pdf}%
%     }\hfill
%     \subfloat[VCA QoE Inference \label{fig:vca_qoe_time_pareto}]{%
%     \includegraphics[width=0.33\textwidth]{figures/vca_qoe_pareto.pdf}%
%     }
%     \caption{\todo{Choose one between this and the former one.}\change{We use Pareto front to determine the candidates of 
%     models.} }
%     \label{fig:pareto_each}
% \end{figure}

\begin{figure}[!t]
    \centering
    \includegraphics[width=0.8\columnwidth]{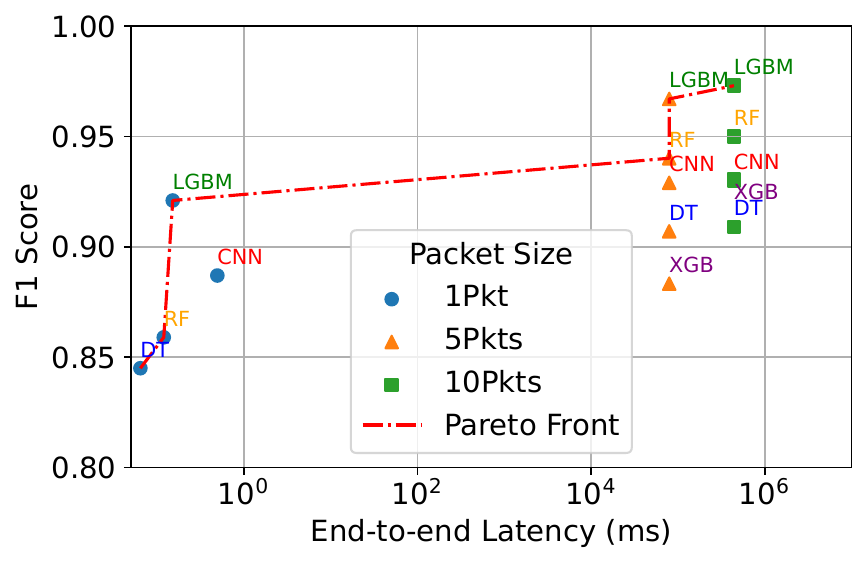}
    \caption{\change{The F1 score vs. end-to-end latency (in ms) across 5 models 
    for the \textbf{service recognition} dataset. We use the Pareto Front to determine 
    the placement of models in \sysname.}}
    \label{fig:pareto_all}
\end{figure}

\noindent \textbf{How to select and place models in \sysname?} Determining 
the placement of models within \sysname involves striking a balance between 
speed and performance. \change{The process follows two key steps: (1) \textit{model 
training and profiling}: we begin by training models with various architectures 
and context lengths (\eg, 1 to 20 packets). Each model is then profiled based 
on its F1 score and end-to-end latency. (2) \textit{Pareto Front selection}: from these 
profiles, we select models that lie on the Pareto Front, which represents 
the best trade-off between performance (F1 score) and latency, as shown in 
Fig.~\ref{fig:pareto_all}. Models on this front achieve the optimal balance, 
avoiding scenarios where improvements in one metric significantly degrade 
the other.} \rchange{This profiling is a one-time offline cost, not online serving
overhead. In our pipeline, searching 5 models across 20 packet depths
(\ie, 100 configurations) takes on the order of hours to about a day depending
on the hardware, and the online system serves only the selected models rather
than the full candidate pool.}  

\change{The selection criterion is as follows: (1) The fastest model -- typically 
with the lowest latency -- is deployed at the entry point to process the initial 
packet of each flow, ensuring rapid early-stage inferences, provided the F1 
score remains within an acceptable range. In Fig.~\ref{fig:pareto_all}, the 
decision tree (DT) with the first packet as input is the fastest model. (2) The 
fast model is usually the most accurate among the first packet models (\eg, 
LightGBM for the service recognition task). 
(3) For the slow model, we determine the number of packets required when the 
F1 score stops improving significantly with additional context (\eg, 10 packets for service recognition). This model, often achieving the highest F1 score,
acts as a slow but precise classifier. Note 
that \sysname may omit the 2nd or 3rd model if there are no significant performance disparities 
that justify the added system cost.}

\subsection{Flow Assignment Algorithm} \label{ss:redirection-methods}
Assigning all flow classification requests to a slow model is not a good idea, 
not only because it harms the latency, but also because 
there is a risk of the slow model making erroneous corrections, which can further
degrade performance. Surprisingly, even if one were to perfectly predict the
correctness of these predictions (akin to having an ``oracle''), this approach
might still negatively affect the overall predictive performance as more requests are assigned to the 
slow model. This counterintuitive outcome is illustrated in 
Figure~\ref{fig:serveflow_teaser}, which necessitates us to find a good assignment 
mechanism to balance the assigned portion with the F1 score.

We essentially need to evaluate the likelihood of predictions made by the fastest or fast
model being ``good enough''.
\change{The objective of assignment strategies is to get a list of thresholds
at each assigning portion,
so that they transfer (a) as many misclassified flows as possible and (b) as 
few correctly classified ones as possible from a faster to a slower model. Ideally, 
this allows the latter model to fix the errors as much as possible, 
while minimizing the overall system cost. }

\noindent \textbf{Universal uncertainty thresholds.} The core of \sysname's efficiency lies in its assignment algorithm, which
discerns the correctness of predictions made by the faster model. This algorithm
is crafted to strategically assign subsets of data based on the uncertainty of
their predictions. Specifically, this uncertainty is quantified using measures
such as least confidence $LC(x) = 1 - \max\limits_{y} P(y|x)$, where $P(y|x)$ is
the probability of the predicted label $y$, or entropy 
$H(x) = -\sum_{i}P(y_i|x) \log P(y_i|x)$, which accounts for the unpredictability inherent in the
probability distribution of the predictions. These metrics, traditionally 
associated with sample selection for labeling during the training phase 
~\cite{liu2023amir,settles2009active}, are repurposed in \sysname for inference in 
fast-slow design.

This assessment is anchored on the use of universal uncertainty thresholds.
Given a validation set $D$ \change{and a model $M$}, it computes the 
uncertainty score for each data point in $D$. \change{The algorithm then 
constructs an ordered list of uncertainty scores and computes the uncertainty 
score at each quantile of the distribution, providing a granular view of 
data uncertainty with their assigned portions (\ie, at each quantile). The
details are in Algorithm~\ref{alg:universal_threshold} in Appendix~\ref{app:implementation-details}.}

\noindent \textbf{Slope-based per-class uncertainty thresholds.} In contrast 
to a universal threshold, \sysname can also employ a more nuanced,
class-specific approach (as described in Algorithm~\ref{alg:per_class_thresholds} in Appendix~\ref{app:implementation-details}).
This approach recognizes that different classes may have varying levels of
prediction certainty and, as such, different thresholds. The algorithm first
segregates the dataset into correct and incorrect predictions per class label.
It then establishes uncertainty thresholds for each class by examining the
quantiles of uncertainty values. The key step involves calculating the
assigned portions for correct and incorrect predictions at each threshold and
using this information to determine the optimal slope for the assignment. This
slope, a ratio of incorrectly assigned predictions to the total assigned,
guides the setting of class-specific thresholds. \change{When we increment 
each assigned portion,} the algorithm 
adjusts the uncertainty threshold for each class, tailoring the assignment
process to improve the system's overall predictive reliability and efficiency.
% \todo{Simplify Alg 2}

\noindent \textbf{Determining the optimal threshold.} \change{With the assigned portions 
and corresponding uncertainty thresholds (whether from the Universal Uncertainty or Per-Class 
Uncertainty algorithm), the next step is to determine the optimal thresholds. One 
approach is to identify the Pareto optimal point, where no further improvement 
can be made in one metric (e.g., F1 score) without compromising the other 
(e.g., latency). This means profiling on validation set $D$ and finding the point at which any further improvement 
in classification accuracy would result in a disproportionate increase in 
latency or resource consumption, and vice versa. Alternatively, a more goal-oriented approach involves setting 
a specific target for either latency or service rate. For instance, if it
requires a particular latency bound or a guaranteed service rate, \sysname can 
adjust the uncertainty thresholds to match this target. In practice, this 
means selecting the appropriate portion of traffic (\ie, quantile) from the distribution of uncertainty 
scores.}
    
\change{The choice for an uncertainty approach
depends on the types of models in use (see details in Sec.~\ref{subsec:flow_assign_eval}).} 
The Per-Class Uncertainty method excels with gradient boosting machines 
\change{(\eg, LightGBM, XGBoost) }, which are designed to iteratively 
improve on areas where the model previously made errors~\cite{ke2017lightgbm}. 
This design can lead to a more nuanced understanding of the data and, 
consequently, a more class-specific distribution of prediction confidence. 
\change{The Universal Uncertainty approach works better for models like decision trees, where
the uncertainty scores are less class dependent.} \sysname configures its assignment 
strategy, selecting the better approach for each stage based on the models.

\section{Model-Serving for Fast-Slow Models}\label{sec:implementation}

In this section, we introduce the design of supporting components required to realize
\sysname.
% (its implementation is detailed in Appendix~\ref{sec:implementation}).

\subsection{Traffic Extraction and Management}
In order to make machine learning tasks possible, traffic extraction is the
first step. At the very front of \sysname, we extract the full nPrint
representations~\cite{holland2021new} in a \textit{real-time} manner, 
\change{which is challenging to achieve at high network throughput in prior 
work~\cite{lotfollahi2020deep,jiang2023acdc,jiang2023netdiffusion}}. nPrint is a
standard data representation for network traffic. It translates all packet
header field bits into a unified representation. This method, even with a
single-packet nPrint, has been demonstrated to provide sufficient information
for robust classification performance, outperforming other flow statistics-based
methods that require multiple packets and often yield less
accuracy~\cite{jiang2023acdc,jiang2023netdiffusion}. \change{This enables our
fastest and fast models on just the 1st packet possible. } However, such full
header bits in raw are challenging to extract in real time. To address this, \sysname
harnesses PF\_RING APIs~\cite{pfring2023}, optimizing the nPrint extraction process for real-time
application. This implementation not only facilitates rapid packet capture and
feature extraction but also maintains the ability to track flows via their
five-tuple keys. Consequently, \sysname is capable of generating
high-dimensional (often thousands per flow) feature vectors instantaneously,
thereby fully utilizing nPrint's robust packet-level representations for
efficient and accurate traffic classification.

We use a sophisticated queueing management mechanism
(implemented using Pulsar APIs~\cite{pulsar2024}) for dynamic traffic flow state
management, which is engineered to handle asynchronous traffic
flows. Each time a new flow request comes to the network interface, we extract
the features and push them to the right Queue. Flow IDs across
Queues and models are also managed. In particular, as shown 
in Figure~\ref{fig:serveflow}, Queue 2 is designed with a long queue to accommodate
the fast model inference and flow of network traffic, ensuring that data is
neither lost nor excessively delayed. This queue operates on a conditional
processing protocol: it primarily holds incoming packet features until the fast
model sends a request signal. However,
it is not uncommon for packet features to accumulate faster than the fast model
can process them, especially during peak traffic periods. It also implements 
a discard policy. If packet features remain unrequested beyond a
predetermined time threshold, they are proactively purged from the Queues.
% \todo{we also talk about the buffer management in section 3.1. However, we haven't talked about buffer size and how our data retention policy impacts accuracy.}

\begin{figure}[t!]
    \centering
    \includegraphics[width=\columnwidth]{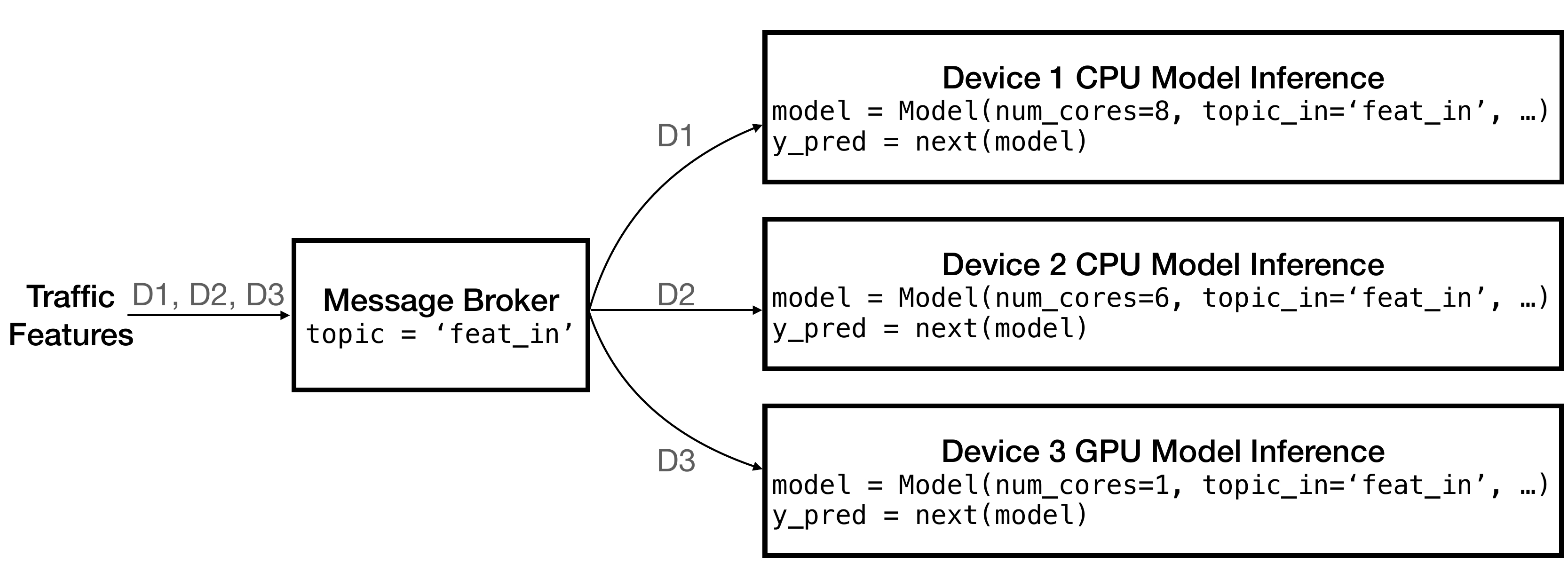}
    \caption{\sysname supports parallelism across heterogeneous hardware.}
    \label{fig:parallelism}
\end{figure}

\subsection{Parallelism Across Hardware Types}
\sysname's queueing mechanism opens up possibilities to enable multiple workers to
concurrently perform inference on the same data stream, which is crucial for
systems handling massive and dynamic traffic volumes. \rchange{The intended
multi-worker design is brokered and keyed by flow ID, not shared-memory state.
First-packet features, later-packet features, and escalation requests are
published as keyed streams; slower workers join these streams by flow ID. For
example, an ISP or an enterprise could benefit by analyzing traffic at high
flow-arrival rates using a combination of existing hardware.} The system's design
abstracts the interfaces to tap into the processing powers of diverse hardware
configurations (different numbers of CPU and GPU workers), thereby optimizing
resource allocation and catering to a range
of computational needs.

Central to \sysname's infrastructure are abstraction layers that seamlessly
distribute tasks across various processing units such as CPUs and GPUs. 
This flexible architecture ensures that each processing unit can serve
requests effectively and efficiently, contributing to the system's overall
performance and responsiveness. 

The advantages of this queueing design are manifold, particularly in
environments characterized by variable workloads and the availability of
computational resources. It confers upon the system an enhanced degree of
flexibility and adaptability, as it allows for dynamic processing by multiple
consumers tapping into the same data stream. This design also affords users the
ability to easily scale the system up or down by adding or removing consumer
nodes, adapting to real-time processing demands without the need for
manual message rescheduling. Such a parallel processing arrangement is essential
for the rapid and efficient management of large volumes, markedly
diminishing latency and upholding scalability.
\rchange{This mechanism also clarifies the boundary relative to the serving
substrate: the broker handles partitioning, flow-ID routing, and durable stream
state, while model workers consume keyed feature streams and return predictions.
Beyond one server, broker throughput, partition balance, and feature transfer set
the main limits.}

Depicted in Figure~\ref{fig:parallelism}, \sysname supports parallelism across heterogeneous hardware
platforms, enabling each consumer to process a segment of the data independently.
The aggregated results from these nodes culminate in the final output, a
demonstration of \sysname's design that not only maximizes the
utilization of resources but also adeptly accommodates high throughput. 

Moreover, the flexibility of \sysname extends to its fast-slow model
architecture, where each model can be assigned a variable number of consumers
according to the computational resources designated by the user at each stage.
This allows for granular customization of resource allocation, ensuring that
each stage is sufficiently resourced to meet the demands.

\subsection{Model Crafting Pipeline}
To support \sysname's fast-slow architecture, an automated training pipeline is
integral. This process starts with loading a training set, where the pipeline
generates nPrints for each flow, converting network data into a
model-friendly format. The key to this stage is the removal of redundant information
to optimize feature quality: it eliminates columns with uniform values
and those duplicating others. This step ensures the retention of only unique,
informative features. It supports 
a variety of training frameworks (\eg, PyTorch~\cite{paszke2019pytorch}, 
TensorFlow~\cite{tensorflow2015-whitepaper}, Scikit learn~\cite{scikit-learn}, 
LightGBM~\cite{ke2017lightgbm}, XGBoost~\cite{xgboost}, etc.) for users to build 
highly customized models \change{with extensible model architectures}.

The next phase involves training a pool of models using these refined features
from the initial packets. Each model undergoes an evaluation of its
inference speed and performance on a validation set. This comprehensive
assessment allows for the selection of models best suited for different roles
within \sysname -- fastest, fast, and slow. For inferences based on a
single-packet flow, the fastest model is chosen for its lowest latency, while 
the fast model is decided through trial-and-error iterations, which ensure 
the least flow portion to assign while achieving high F1 scores. The slow
model is the one with the best F1-score among all. These selected models,
together with their input feature formats (\ie, subscriptions), are transformed
into ONNX format~\cite{onnxruntime}, which enables inter-operability across
different frameworks and tools for efficient model optimization and serving.
They are later deployed in the fast-slow architecture.
\rchange{This training and profiling cost is paid offline. In practice, the
automated search over 5 model families and 20 packet depths takes on the order
of hours to about a day, depending on the hardware. The online serving path
loads only the selected fastest, fast, and slow models.}
\rchange{Appendix~\ref{app:implementation-details} provides the implementation
details for packet capture, feature subscriptions, model configuration, and
serving.}

\section{Evaluation}\label{sec:evaluation}
In this section, we holistically evaluate the performance of 
\sysname across three datasets collected in the real world.
\subsection{Evaluation Settings}\label{subsec:setting}
We detail our tasks, datasets, and performance indicators, consisting of 3
unique settings. Appendix~\ref{app:evaluation-setting-details} provides the testbed configuration and
baseline implementation details.

\paragraph{Service Recognition.} Service recognition plays a vital 
role in tasks such as resource allocation and Quality of Service 
(QoS) assurance~\cite{piet2023ggfast,guthula2023netfound}. In the 
dataset detailed in Appendix~\ref{app:data}, 
the focus is on categorizing network flows across 4 macro services 
encompassing 11 distinct classes of applications. This compilation 
includes a total of 23,487 flows, demonstrating the diversity and 
complexity of network traffic involved in classifying services for 
enhanced network management and service delivery.
\rchange{\underline{\textit{Stress test.}} To stress test the multi-worker
pipeline with a more diverse dataset, we also use CipherSpectrum, a 41-class
TLS 1.3 service recognition dataset~\cite{wickramasinghe2025sok}, scaled to 192 workers in Sec.~\ref{subsec:large-scale-testing}.}

\paragraph{Device Identification.} Identifying IoT devices is a critical step in 
mitigating attacks by isolating these devices through communication restrictions 
from firewalls or gateways~\cite{salman2022machine,aksoy2019automated}.
We leverage the open-source dataset from AMIR~\cite{liu2023amir}, which features 
a diverse collection of 18 in-home devices and encompasses 50,017 unique flows, 
as documented in Appendix~\ref{app:data}.
A notable characteristic of this dataset is the prevalence of short-lived flows, 
which informs our decision to utilize 3 packets for analysis in the slow model.

\paragraph{QoE Measurement.}
QoE measurement on encrypted traffic is an important piece
for the management of video streaming applications~\cite{mangla2018emimic,mangla2018quicqoe,bronzino2019inferring} or video 
conferencing apps (VCA)~\cite{sharma2023estimating,macmillan2021measuring}. 
We use the in-lab datasets from~\cite{sharma2023estimating}
for per-second estimates of key VCA QoE metrics such as frame rate. 
As detailed in Appendix~\ref{app:data}, it has
a total of 36,928 seconds of data across Meet, Webex, and Teams 
to make inferences on.
The assessment categorizes target frame rates into 11 tiers, 
incrementing by 3 frames per second up to a threshold beyond 30 
fps. This task diverges from the previous two by 
necessitating continuous inference by \sysname, requiring it 
to monitor flows and update inferences from second to second.
\paragraph{\rchange{Performance Indicators.}}
\rchange{Latency and throughput are tightly coupled: waiting for
more packets both delays the decision and keeps per-flow state alive longer, which lowers service rate under load.}
\rchange{Thus, we report end-to-end serving latency, measured from a flow's
first observable packet to the returned label. This metric indicates whether
downstream applications can still act while the flow is active. We use new
flows/s as the main metric because \sysname is a flow-level serving system
rather than a wire-speed packet-processing engine. This is a conservative
workload measure: each new flow can imply concurrent older flows and multiple
packets per flow. We also report F1 to ensure that these optimized
configurations preserve accuracy.}

\subsection{System Performance}

In this section, we evaluate the performance of \sysname against 
\change{the following software baselines\footnote{We exclude SmartNIC-based
systems (\eg, Homunculus~\cite{swamy2023homunculus} and
LEO~\cite{jafri2024leo}) because they require specialized hardware; \sysname's
flow-collection optimization is orthogonal to their designs.}, summarized in
Table~\ref{tab:traffic_analysis_systems}.}

\begin{table}[t]
  \centering
  \small
  \setlength{\tabcolsep}{0.6pt}
  \begin{tabular*}{\columnwidth}{@{}l@{\extracolsep{\fill}}c c c c@{}}
      \toprule
      \textbf{Method} & \shortstack{\textbf{Packet}\\\textbf{Depth}} & \textbf{Feature} & \textbf{Model(s)} & \textbf{Queue} \\
      \midrule
      LEXNet~\cite{fauvel2023lightweight} & N & Pkt Size \& Direction & LEXNet (CNN) & No \\
      FastTraffic~\cite{xu2023fasttraffic} & N & N-gram of Raw Pkts & 3-layer MLP & No \\
      Best Effort & N & nPrint & LGBM & No \\
      Queueing & N & nPrint & LGBM & Yes \\
      \sysname & 1+N & nPrint & DT+LGBM+LGBM & Yes \\
      \bottomrule
  \end{tabular*}
  \caption{Comparison of evaluated traffic analysis frameworks. N is the best 
  packet depth for an application (10 for service recognition, 3 for device 
  identification, 20 for QoE inference.)}
  \label{tab:traffic_analysis_systems}
  \vspace{-2.5em}
\end{table}

\begin{itemize}[leftmargin=*, itemsep=0.1em, topsep=0.2em]
  \item \textbf{LEXNet}~\cite{fauvel2023lightweight} is a lightweight CNN
  traffic classifier that uses only packet size and direction. We use the same
  packet depth as the other baselines for a fair comparison.
  \item \textbf{FastTraffic}~\cite{xu2023fasttraffic} is a packet-level neural
  classifier based on truncated packet bytes and N-gram embeddings. We adapt it
  to per-flow inference to avoid repeated per-packet computation.
  \item \textbf{Best Effort} extracts nPrint features with PF\_RING and runs the
  ONNX model immediately after the best-N packet features are ready, all within
  the same C++ thread.
  \item \textbf{Queueing} uses the same feature queue and FIFO processing as
  \sysname, but removes the fast-slow architecture.
\end{itemize}

Appendix~\ref{app:evaluation-setting-details} gives the evaluation setting details.

The baseline approaches are configured to perform model inference using the best N packets of each flow (N identified 
with the most performant model).
\sysname is configured to achieve its best possible performance-accuracy tradeoff.
We evaluate this tradeoff in the remaining sections of the evaluation. \sysname is able to 
achieve \textit{a similar F1 score} compared to running \textit{all} requests to slow model.
Following the \change{model selection and placement procedures described in Sec.~\ref{sec:fast-slow-arch}}, \sysname uses 
a decision tree on the first packet as the fastest model, a LightGBM model on the 
first packet as the fast model, and another LightGBM model on the ten first packets 
as the slow model.
Further, \sysname assigns requests from the fastest to the fast model according 
to the Uncertainty approach and from the fast to the slow model using the 
Per-Class Uncertainty approach.
% To draw a fair comparison, we limit \sysname and baselines to use a single-core.
\textit{In the interest of space, we focus on Service Recognition.
We observed similar results for the other tasks (see Appendix~\ref{app:perf}).}

\begin{figure}[!t]
  \centering
  \subfloat[Single-consumer\\ service rate. \label{fig:trp_comp}]{%
    \includegraphics[width=0.48\columnwidth]{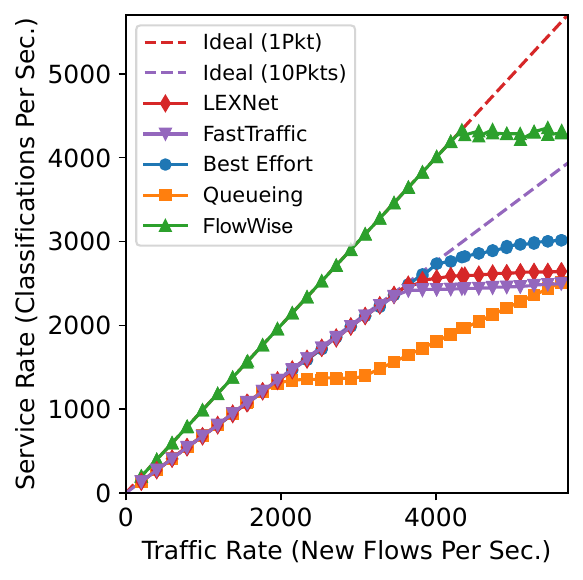}%
  }
  \subfloat[Average end-to-end \\ latency. \label{fig:se2e_t_comp}]{%
    \includegraphics[width=0.48\columnwidth]{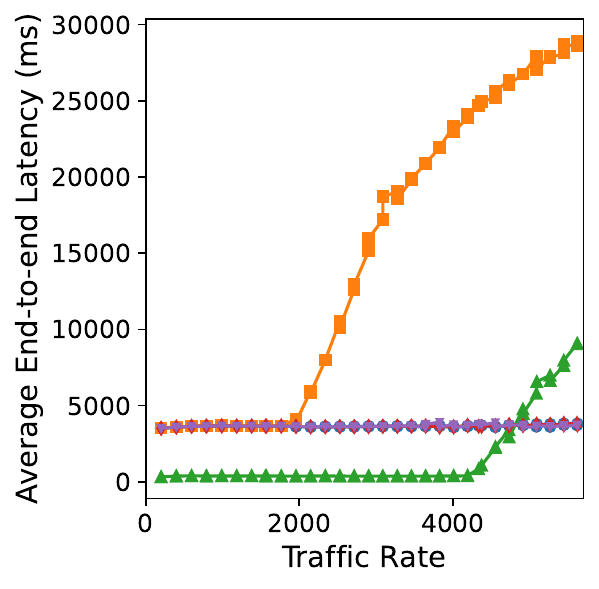}%
  }
  % \subfloat[Miss rate (All). \label{fig:lr_comp}]{%
  %   \includegraphics[width=0.23\columnwidth]{figures/lossrate_srec.pdf}%
  % }
  \\
  \subfloat[Miss rate (10Pkts). \label{fig:long_lr_comp}]{%
    \includegraphics[width=0.47\columnwidth]{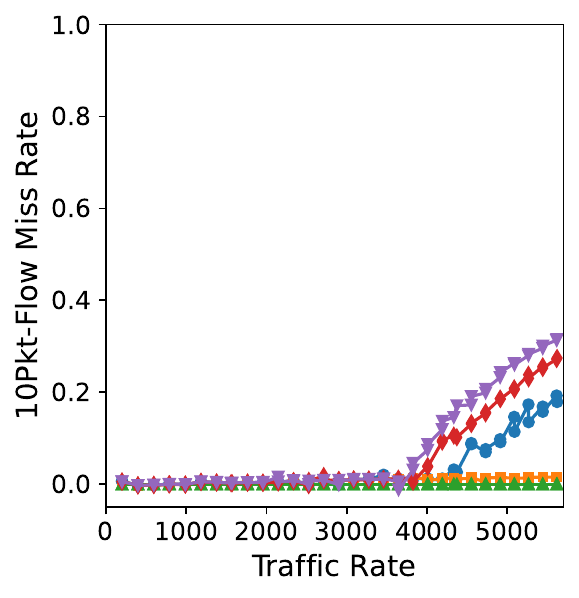}%
  }
  \subfloat[Max service rate vs. F1 \label{fig:service_rate_f1}]{%
  \includegraphics[width=0.48\columnwidth]{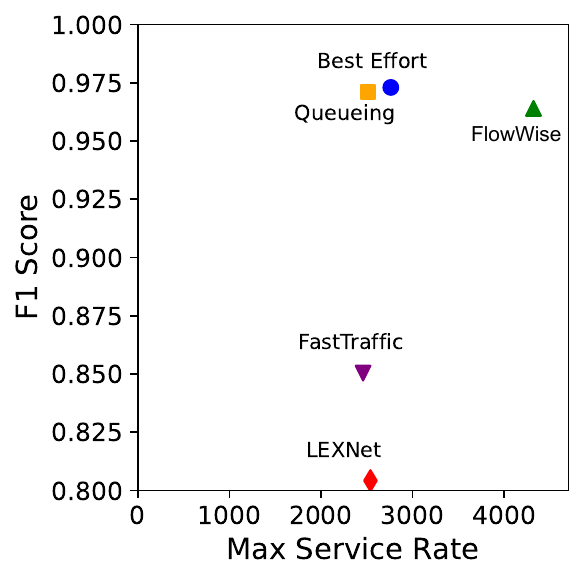}%
}
  \caption{Performance of different approaches for the \textit{service recognition} task (other 
  tasks are in Appendix~\ref{app:perf}). We compare \sysname against 
  four baselines, all considering 10-packet inputs. \change{The F1 score 
  is computed on the non-missed flows.}}
  \label{fig:base_comp}
\end{figure}

\begin{figure}[t!]
  \centering
  \subfloat[CDF of end-to-end latency. \label{fig:E2E_CDF}]{%
    \includegraphics[width=0.85\columnwidth]{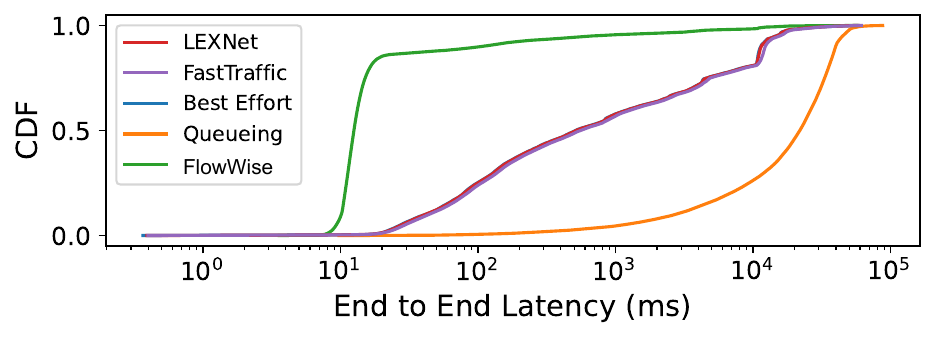}%
  }\hfill
  \subfloat[Latency introduced at different stages. \label{fig:stage_breakdown}]{%
    \includegraphics[width=0.85\columnwidth]{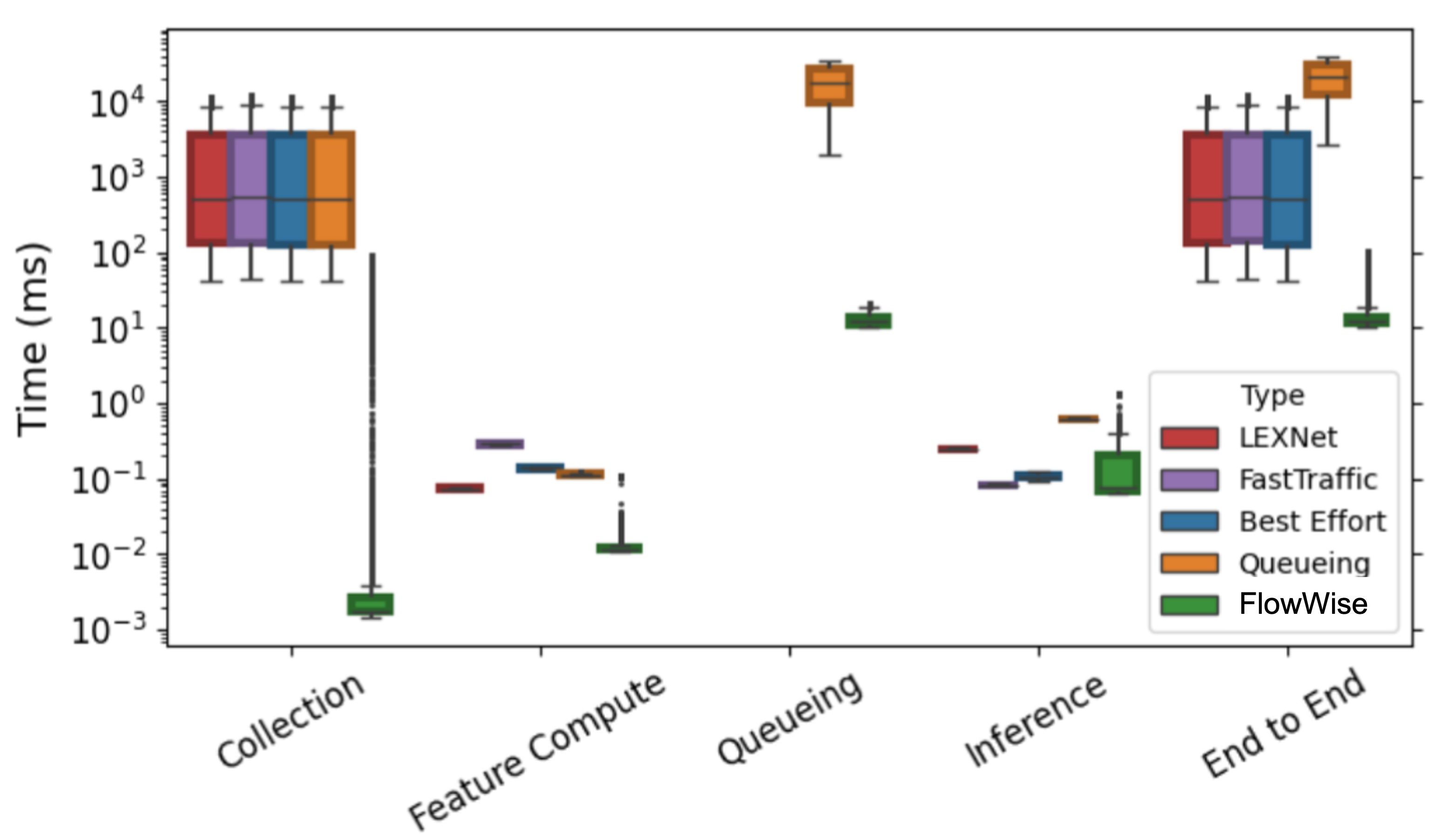}%
  }

  \caption{End-to-end latency analysis of 4k flows per second on a single consumer
  (\ie, 1 CPU core) for \textit{service recognition}: (a) The
  CDF of single-consumer latency demonstrates that \sysname can
  accelerate 76\% of flows to millisecond-level service. (b) A
  breakdown of latency at different stages illustrates the reduction
  of packet waiting time, showing the efficiency of \sysname in
  minimizing overall latency. }
  \label{fig:latency_breakdown}
% \vspace{-1em}
\end{figure}

\paragraph{Service rate, end-to-end latency, miss rate, and F1 Score.}
We define \textit{service rate} as the number of predicted flows per second; 
\textit{end-to-end latency} as the total time spent for a flow between 
its first packet is produced and a satisfactory prediction is issued, 
and \textit{miss rate} as the percentage of flows without a prediction 
issued by the model.
Figure~\ref{fig:base_comp} presents (a) the single-consumer service rate, 
(b) the average end-to-end latency, 
and (c) \change{ten-packet} flow miss rate for approaches as a function 
of the network traffic rate, i.e., the number of new flows arriving per 
second. \change{(d) the F1 score (of classified flows) and the maximum 
service rate each method can achieve.}
In Fig.~\ref{fig:trp_comp} we also include two lines, i.e., 
Ideal (1Pkt) and Ideal (10Pkts), to indicate the effective traffic 
rate when the approaches have to wait for one or ten packets before 
inference, respectively. \change{Ideal (10Pkts) is lower because 31\%
of the flows do not have a length of 10 packets in service recognition.}

From the figures, we observe that \sysname achieves better service rate
than the other approaches, keeping up with the traffic rate for 
a single packet up until around 4.3k flows per second.
We also observe that the average latency for \sysname is consistently 
several orders of magnitude smaller than the other baseline approaches 
before resource saturation. After 4.3k flows per second, the queue becomes 
congested and it starts to increase the latency. And it achieves almost 0.0\% 
miss rate because of internal queues, as well as precise-number flow 
tracking because it does not have to wait for 10 packets for 
a flow. \change{Plus, \sysname has a similar F1 score (0.964) compared to 
Best Effort (0.973) on classified flows.}

In contrast, Queueing saturates at around 2.5k flows per second 
while \change{LEXNet, FastTraffic, and Best Effort keep the latency 
consistent (Fig.~\ref{fig:se2e_t_comp}) as they start to miss flows when 
they max their service rate (Fig.~\ref{fig:long_lr_comp}). For the F1 scores,
both LEXNet and FastTraffic do not perform as well as the other methods using 
LightGBM.} 
These results show the evidence of \sysname's capacity to improve the overall performance by
serving most of the flows confidently with the faster models, which only 
require a single packet as input. And Figure~\ref{fig:latency_breakdown} further
illustrates this.

\paragraph{Latency breakdown.} 
For this analysis, we fix the traffic rate at 4k flows per second and plot the CDF for 
end-to-end latency across inference instances (Fig.~\ref{fig:E2E_CDF}).
We observe that \sysname is capable of serving about 76\% of the flows within 16~ms.
In Fig.~\ref{fig:stage_breakdown}, we also analyze the latency breakdown at each serving stage.
We note that the collection delay for \sysname is about \change{six to eight} orders of magnitude 
faster than the baseline approaches. When compared to the Best Effort, \change{LEXNet, and 
FastTraffic} approaches, the added queueing delay and the marginal increase in the 
inference delay (due to running multiple models for some flows) are quickly offset by 
these savings in the collection stage.
All these results together showcase the performance benefits of the proposed 
fast-slow architecture for network traffic analysis. Next, we evaluate how \sysname 
enables exploring the tradeoff between performance and accuracy.

\subsection{Flow Assignment Algorithm Performance}\label{subsec:flow_assign_eval}

The efficiency and accuracy of \sysname depend on the assignment approach.
We compare four approaches: Oracle, Random, Uncertainty, and Per-class
Uncertainty.
\begin{itemize}[leftmargin=*, itemsep=0.1em, topsep=0.2em]
  \item \textbf{Oracle.} This idealized approach assigns flows using
  ground-truth knowledge of whether each model prediction is correct.
  \item \textbf{Random.} This approach assigns flows by random sampling.
  \item \textbf{Uncertainty.} This approach uses
  Algorithm~\ref{alg:universal_threshold} in Appendix~\ref{app:implementation-details}.
  \item \textbf{Per-Class Uncertainty.} This approach uses
  Algorithm~\ref{alg:per_class_thresholds} in Appendix~\ref{app:implementation-details}.
\end{itemize}

\begin{figure}[!t]
  \centering
  \subfloat[Service recognition \label{fig:serv_rec_portion_1}]{%
    \includegraphics[width=0.33\columnwidth]{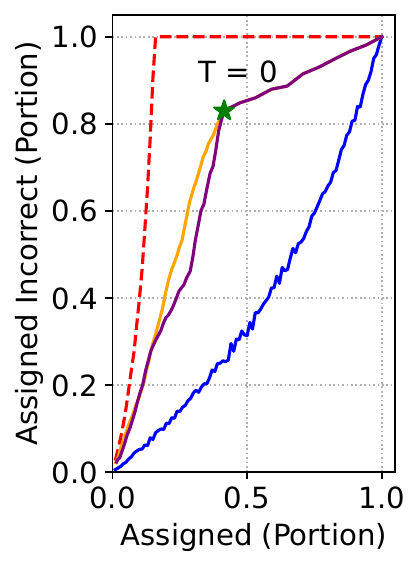}%
  }
  \subfloat[Device id. \label{fig:dev_id_portion_1}]{%
    \includegraphics[width=0.33\columnwidth]{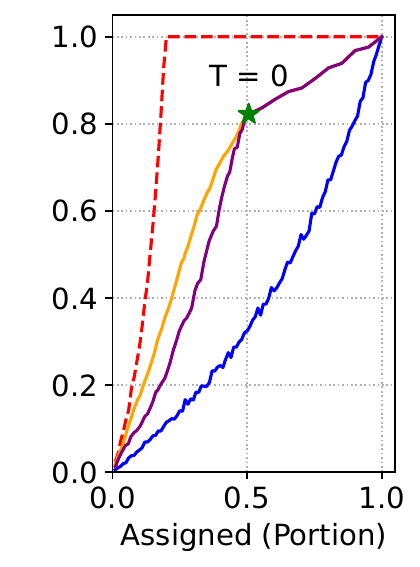}%
  }
  \subfloat[QoE inference \label{fig:vca_qoe_portion_1}]{%
    \includegraphics[width=0.33\columnwidth]{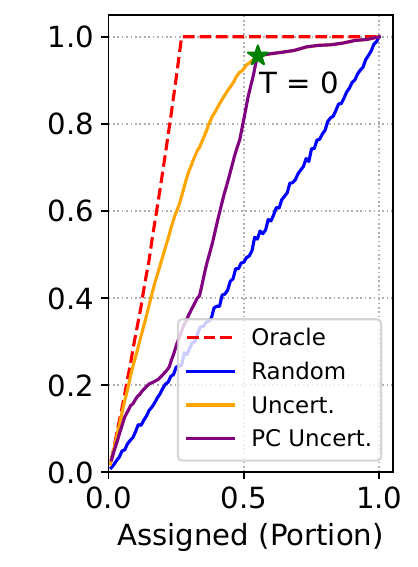}%
  }
  \caption{Assigned portion of requests \change{(\#assigned flows / \#all flows)} vs. 
  assigned incorrect portion \change{(\#assigned incorrect predictions / \#all 
  incorrect predictions)} using different approaches on the fastest model. T: Uncertainty threshold for the Uncertainty and Per-class Uncertainty 
  approaches. }
  \label{fig:portion_1}
\vspace{-0.8em}
\end{figure}
\paragraph{Assigned vs. assigned incorrect portion.}
The goal of assignment approaches is to assign (a) as many incorrectly
classified flows as possible, and (b) as few correctly classified ones as possible, from a
\textit{faster less accurate} model to a \textit{slower more accurate} one. In the
ideal case, this maximizes the opportunity for the latter model to correct the
mistaken classifications (improving accuracy) while minimizing the overall
system end-to-end latency. 

We evaluate how well assignment approaches attain
this goal. Figure~\ref{fig:portion_1} presents the percentage of the incorrect
classifications that are assigned as a function of the percentage of overall
requests that are assigned by each approach. Our proposed Uncertainty and 
Per-class Uncertainty methods provide a good trade-off between the Oracle 
and random assignment. Both methods successfully assign 82-95\% of the 
incorrect classifications while only assigning about 50\% of the total 
requests. For the decision tree model, the Uncertainty approach 
converges to the zero threshold more quickly.

\begin{figure}[!t]
    \centering
    \subfloat[After fastest to fast model. \label{fig:serv_rec_f2f} ]{%
      \includegraphics[width=0.75\columnwidth]{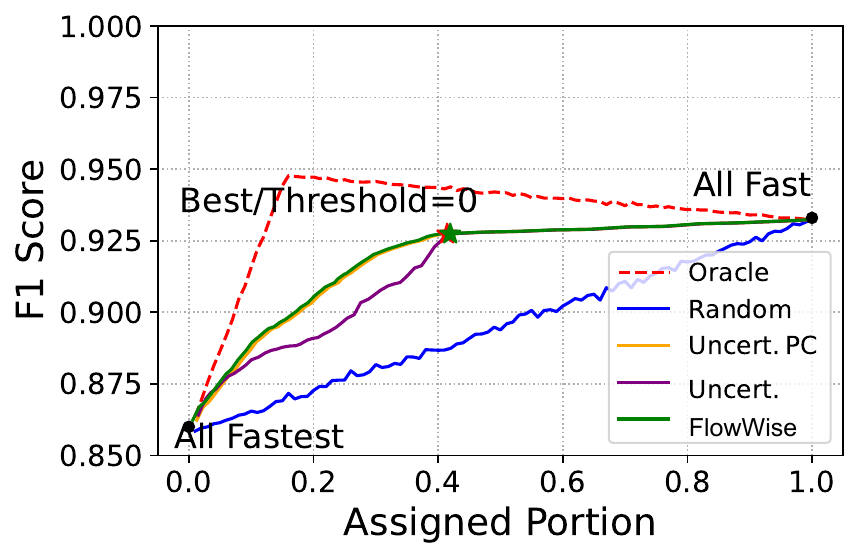}%
    }
    % \subfloat[Fast model to slow model. \label{fig:serv_rec_f2s}]{%
    %   \includegraphics[width=0.33\textwidth]{figures/service_recognition_pVsf1_f2s.pdf}%
    % }\hfill
    \hspace{0.1\columnwidth}
    \subfloat[After fast to slow model. \label{fig:serv_rec_e2e} ]{%
      \includegraphics[width=0.75\columnwidth]{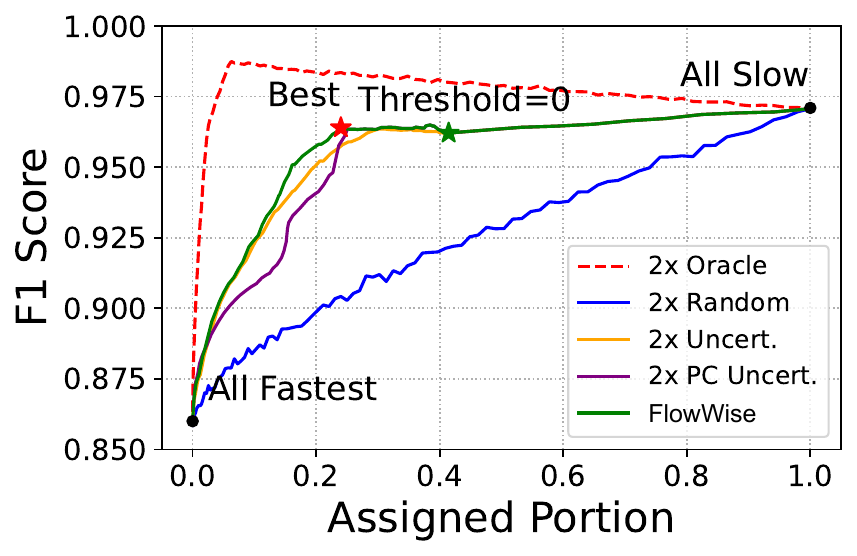}%
    }
    \caption{Assigned portion of requests vs. F1 score for \textit{service recognition}.
    \change{Based on models, }\sysname is best configured to use Uncertainty for the first assignment,
    and Per-Class Uncertainty for the second.}
    \label{fig:portion_f1}
  \end{figure}

\paragraph{Assigned portion vs. F1 score.}
Given the ability of assignment approaches to discern between correct and
incorrect, next we evaluate their impact on the accuracy of the system. We
analyze the F1 score after each stage, i.e, after fastest to
fast (Fig.~\ref{fig:serv_rec_f2f}) and after fast to slow (Fig.~\ref{fig:serv_rec_e2e}).

The first insight we draw from these two figures is that: even for Oracle, 
the best approach to achieve a high F1 score is not to assign everything to 
a slow model. \change{The slower model may experience mismatches with faster 
model due to undertraining or overfitting to specific samples, or differences 
in architecture and parameterization}, 
and make correct predictions from faster models faulty. Instead, the most 
ideal assigned portions are 17\% and 7\%, respectively.

Uncertainty-based methods show a better balance between Oracle and Random.
Essentially, Uncertainty works better for the first stage for decision trees,
while Per-Class Uncertainty arrives at a saturated F1 score 6\%
assignments earlier.

Since \sysname configures with the best assignment approach based 
on different types of models (which will be presented later), it is able 
to arrive at the best trade-off point earlier than other non-Oracle baselines: 
around 24\% assigned flow to get a 0.962 F1 score on all the flows.

\paragraph{Assignment effectiveness vs. model.} 
We further investigate that within \sysname 
how different models respond to various assignment strategies. We use
the Normalized Area Under Curve (AUC) metric to measure the effectiveness of
them by comparing the improvement in F1 score relative to the
fastest model, normalized against Oracle's absolute AUC. A higher normalized
AUC indicates a better balance between speed and F1 score improvement. 

The findings, detailed in Table~\ref{tab:nauc_table}, highlight differences in
assignment effectiveness when altering the fastest and fast models in \sysname. 
In scenarios where the fastest model varies (Table~\ref{subtab:nauc_vary_fastest_model}), 
Uncertainty-based assignment shows greater benefits
for Decision Trees and Random Forest. Conversely, Per-Class Uncertainty
significantly improves assignment for XGBoost models, demonstrating a notable
advantage (0.874 vs. 0.674 for Uncertainty). When the fast model varies, with
Decision Tree as the constant fastest model (Table~\ref{subtab:nauc_vary_fast_model}), 
Uncertainty generally fares better for models other than
boosting machines. However, Per-Class Uncertainty is particularly effective when
LightGBM is the fast model.

This investigation shows that the variability in confidence distribution across
classes, owing to how each model type inherently processes and learns from the
data, is the crucial factor. For models like DT, RF, and CNN, where this
variability is lower or managed in a uniform way~\cite{wehenkel1996uncertainty}, 
a universal Uncertainty threshold suffices. For boosting machines, where
variability is high and class-specific nuances are
significant~\cite{ke2017lightgbm}, Per-Class Uncertainty thresholds become
essential to accurately gauge prediction reliability.

\begin{table}[!t]
  \centering
  % \small

  \begin{subtable}{0.96\columnwidth}
      \centering
      \footnotesize
      \begin{tabular*}{\columnwidth}{@{\extracolsep{\fill}}l l l c c c@{}}
          \toprule
          \textbf{Fastest} & \textbf{Fast} & \textbf{Slow} & \textbf{Rand.} & \textbf{Uncert.} & \textbf{PC Uncert.} \\
          \midrule
          DT & \multirow{4}{*}{LGBM} & \multirow{4}{*}{LGBM} & 0.514 & \cellcolor[HTML]{b7b7b7}\textbf{0.789} & 0.717 \\
          RF &  &  & 0.535 & \cellcolor[HTML]{b7b7b7}\textbf{0.834} & 0.789 \\
          XGB & &  & 0.591 & 0.674 & \cellcolor[HTML]{b7b7b7}\textbf{0.874} \\
          CNN & &  & 0.468 & \cellcolor[HTML]{b7b7b7}\textbf{0.721} & 0.704 \\
          \bottomrule
      \end{tabular*}
      \caption{Vary fastest model.}
      \label{subtab:nauc_vary_fastest_model}
  \end{subtable}
  \hspace{0.02\columnwidth} % Space between the two tables
  \begin{subtable}{0.96\columnwidth}
      \centering
      \footnotesize
      \begin{tabular*}{\columnwidth}{@{\extracolsep{\fill}}l l l c c c@{}}
          \toprule
          \textbf{Fastest} & \textbf{Fast} & \textbf{Slow} & \textbf{Rand.} & \textbf{Uncert.} & \textbf{PC Uncert.} \\
          \midrule
          \multirow{4}{*}{DT}  & RF & \multirow{4}{*}{LGBM} & 0.664 & \cellcolor[HTML]{b7b7b7}\textbf{0.793} & 0.760 \\
           & LGBM &  & 0.807 & 0.817 & \cellcolor[HTML]{b7b7b7}\textbf{0.838} \\
            & XGB &  & 0.459 & 0.501 & \cellcolor[HTML]{b7b7b7}\textbf{0.771} \\
            & CNN &  & 0.695 & \cellcolor[HTML]{b7b7b7}\textbf{0.789} & 0.784 \\
          \bottomrule
      \end{tabular*}
      \caption{Vary fast model.} 
      \label{subtab:nauc_vary_fast_model}
  \end{subtable}
  \caption{Comparison of Normalized AUC (assigned portion vs. improved F1 score normalized on the Oracle results) across different assignment methods. We vary models in a fastest-fast-slow (DT 1 pkt + LGBM 1 pkt + LGBM 10 pkts) architecture. Highlighted cells indicate the best performance in each. This is evaluated using \textit{service recognition} dataset, but similar trends are observed with other datasets.}
  \label{tab:nauc_table}
\end{table}

\subsection{Microbenchmarks}

\begin{figure}
  \centering
  \includegraphics[width=0.85\columnwidth]{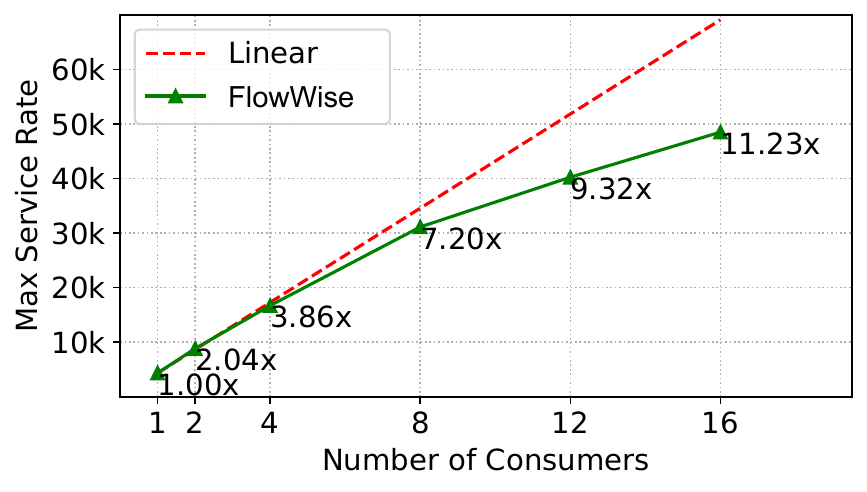}
  \caption{\sysname service rate with the \# of consumers.}
  \label{fig:scalability}
% \vspace{-0.8em}
\end{figure}

\paragraph{Service rate scalability.}
We evaluated how \sysname performance scales as more computational resources are
made available to it. We determine the maximum service rate that \sysname can
achieve as we vary the number and the hardware type of consumers used for
model inference and report results in Figure~\ref{fig:scalability}. As 
expected, the throughput grows sublinearly due to
the additional communication overhead. Nevertheless, by using only 16 consumers
(only CPU or mixed) on a single server, \sysname is already able to keep up with traffic at a rate
of over 48.5k \textit{new} flows per second.
This rate is comparable to reported flow-arrival rates in city-backbone
traces~\cite{CAIDABackbone2024}, and is more than 3x the service rate of the
most recent commodity hardware traffic analysis solution~\cite{jiang2023acdc}.

\begin{table}
  \centering
  \small
  \begin{tabular}{crrrrr}
    \toprule
    \textbf{Packet Number} & \textbf{2} & \textbf{4} & \textbf{6} & \textbf{8} & \textbf{10} \\
    \midrule
    \textbf{Final F1 Score} & 0.945 & 0.956 & 0.960 & 0.962 & 0.963 \\
    \textbf{Average Latency} & 132.7 & 188.5 & 323.0 & 358.8 & 372.2 \\
    \textbf{Median Latency} & 11.2 & 11.1 & 11.3 & 11.2 & 11.2 \\
    \textbf{Max Service Rate} & 4285 & 4210 & 4234 & 4226 & 4319 \\
    \bottomrule
  \end{tabular}
  \caption{\sysname performance when varying the number of packets on the slow model,
  for the \textit{service recognition} task. The unit of latency is ms, and the 
  unit of service rate is classified flows per second.}
  \label{table:packet_performance}
\vspace{-1.2em}
\end{table}

\paragraph{\sysname performance vs. packet number.}
We show the performance changes of \sysname when we vary packet number in Table~\ref{table:packet_performance}.
As the number of packets increases from 2 to 10, there is an
improvement in the final F1 score from 0.945 to 0.963. This improvement 
suggests that analyzing more packets allows for more
precise service recognition. However, this accuracy comes at the cost of
increased average latency, which escalates from 132.7 ms to 372.2 ms, reflecting
the additional waiting and processing time required for larger packet numbers. 
The median latency remains stable (even 76\% because that is the
flow portion being served by fast models) across all packet numbers, hovering around 
11.2 ms, indicating that the median processing time is largely
unaffected by the increase in packet count. Throughput, measured as maximum service rate in thousands of flows per second, shows slight fluctuations but
generally remains above 4200 flows per second, peaking at 4319 with 10 packets, which
suggests that \sysname maintains high efficiency and throughput even as the
number of packets increases. This performance evaluation
demonstrates \sysname's capability to balance F1
score, with processing efficiency, as seen in throughput and latency metrics,
across different packet depths.

\subsection{\rchange{Large-Scale Testing}} \label{subsec:large-scale-testing}

\rchange{We further stress-test \sysname on CipherSpectrum~\cite{wickramasinghe2025sok}, a 41-class TLS 1.3
website-traffic dataset. This experiment increases scale in two ways. First, it
increases traffic diversity: it contains more classes than the QoE
workload and covers more diverse encrypted-flow patterns.
Second, it increases serving pressure: we construct a saturated replay workload
that mixes flows from all 41 classes and repeatedly replays them to sustain a
high new-flow arrival stream (1M new flows / s). The replay increases runtime pressure, but does
not change the classification task or create additional classes.}

\rchange{We run the scale-up
experiment on a 2x Intel Xeon Platinum 8480C server. Each processor has
56 physical cores, giving the machine 224 hardware
threads in total. The per-thread CPU capability is comparable to, and slightly
higher than, the 16-core EPYC server used in our earlier experiments
\cite{passmark_xeon_8480c_2026,passmark_epyc_7302p_2026}. This setup allows us
to scale the serving runtime up to \emph{192 workers}. \sysname uses a Decision Tree as the fast model and a GradientBoosting
classifier as the slow model, with threshold \(t=0.10\). }

\rchange{On the saturated CipherSpectrum replay workload, the fast model alone
achieves F1 0.878. Serving every flow with the slow model improves F1 to 0.962,
but makes all flows pay the slow-path cost. \sysname keeps \emph{71.6\%} of
flows on the fast path and escalates only \emph{28.4\%} to the slow path. It
therefore reaches F1 0.955, close to the slow-only F1, while running
\emph{2.28\(\times\)} faster than serving every flow with the slow model. Under
the saturated new-flow stream, \sysname scales to \emph{192 workers} and
sustains \emph{437K new flows/s}. These results show that \sysname preserves its
fast/slow separation on diverse and high-pressure workload.}

\section{Related Work}\label{sec:related}
\noindent \textbf{Network traffic classification.}
Network traffic representation is key in ML for networking, evolving from heuristic-based methods to flow statistics and representation learning. Heuristic methods have lost reliability due to encryption and internet consolidation~\cite{zheng2022mtt, lotfollahi2020deep, boutaba2018mlsurvey, liu2023leaf}. Current approaches use flow statistics (e.g., packet numbers, sizes) via tools like NetFlow and IPFIX~\cite{bernaille2006earlyappid, karagiannis2005blinc, claise2004cisco, claise2008specification}. Advances in machine learning enable learning from raw packets, improving accuracy but reducing efficiency and interpretability. Deep learning, while precise, remains inefficient and opaque~\cite{lotfollahi2020deep, rimmer2017automated, zheng2022mtt, jiang2023acdc, holland2021new}. \sysname leverages nPrint~\cite{holland2021new} for expressiveness, interpretability, and optimized extraction speed.

\noindent \textbf{Traffic classification serving efficiency.}
Several systems have been engineered for efficient traffic classification~\cite{fauvel2023lightweight,xu2023fasttraffic}, 
such as AC-DC~\cite{jiang2023acdc}, which employs a pool of adaptive models 
based on constraints and requirements. Others, like NeuroCuts~\cite{liang2019neural}, 
focus on optimizing decision trees based on performance criteria.
Traffic Refinery specializes in cost-aware model training but lacks automation~\cite{bronzino2021trafficrefinery}.
Solutions like N3IC~\cite{siracusano2022n3ic}, BoS~\cite{yan2024brain} and 
many more~\cite{swamy2023homunculus,swamy2022taurus,jafri2024leo,zhou2023efficient}, 
use specialized hardware to accelerate network traffic classification. 
pForest~\cite{bussegrawitz2019pforest} further explores the trade-off between packet
waiting time and classification accuracy by training a sequence of random forest models
for different phases of a flow on a programmable switch.
\change{BoS~\cite{yan2024brain} uses a P4-programmable switch data plane with
an off-switch module, rather than a SmartNIC redesign. It includes an
escalation path to server-side models when confidence drops, but still requires
waiting for multiple packets before making a prediction.}
The requirement for specialized hardware often implies a lack of complex operations (\eg, floating-point operations, 
dot products, divisions), a small pool of supported ML model architecture (\eg, 
only binary neural networks~\cite{swamy2022taurus}, decision trees, random 
forest~\cite{bussegrawitz2019pforest}, RNNs~\cite{yan2024brain}), a significant 
engineering effort (\eg, building models in a domain-specific language), 
and limited parallelism opportunities.

% Despite their focus on efficiency, these systems seldom address automated feature 
% exploration or have special hardware prerequisites.

A separate vein of research explores low-latency traffic classification. 
Some methods emphasize model simplification for efficiency~\cite{koksal2022markov,
qiu2022traffic, tong2014high}, while others utilize ensemble or tree-based 
algorithms~\cite{devprasad2022context, liu2019adaptive}. However, these 
approaches generally focus on model execution and neglect the optimization 
of feature space and preprocessing. Early flow detection methods aim for latency-aware efficiency. For example, 
early application identification techniques can classify traffic using just 
a few packets~\cite{bernaille2006earlyappid,jiang2023acdc}, an idea also incorporated in 
the design of \sysname.

\rchange{\noindent \textbf{General model serving and speculative execution.}
General-purpose serving systems such as Clipper, InferLine, AlpaServe, INFaaS,
AdaInf, and MArk focus on low-latency prediction, provisioning, scheduling, and
multi-model placement for complete inference requests~\cite{crankshaw2017clipper,
crankshaw2020inferline,li2023alpaserve,romero2021infaas,shubha2023adainf,
zhang2019mark}. Speculative decoding and speculative sampling use a fast draft
model with a slower verifier to reduce autoregressive decoding latency while
preserving the target model distribution~\cite{leviathan2023fast,
chen2023accelerating}. \sysname shares the broad fast-slow pattern, but addresses
a different systems problem: streaming network flows require per-flow state,
packet waiting, flow-ID joins, and uncertainty-based slow-path assignment before
an inference request is complete.}

\section{Conclusion and Future Work}\label{sec:conclusion}
This paper presents \sysname, a
fast-slow model serving architecture for ML-based network traffic analysis on
streaming flows. The key insight is that packet waiting time often dominates
end-to-end latency, while models for the same task can differ substantially in
inference cost and accuracy. \sysname leverages these gaps by issuing early
predictions from lightweight one-packet models and escalating only uncertain
flows to slower models that consume richer multi-packet context. A queue-centric
runtime maintains per-flow state, supports asynchronous feature joining, and
scales inference across heterogeneous CPU/GPU resources. 

More broadly, \sysname suggests a general design pattern for latency-critical
inference on streaming inputs: treat observation time, feature computation, and
inference as one end-to-end budget, and allocate additional packets and compute
only when extra evidence is likely to change the decision. This perspective can
shape a new class of online  control loops, where early
classifications trigger immediate actions (e.g., prioritization, isolation, or
mitigation) and later stages refine decisions. It also
has the potential to be applied beyond traffic analysis: similar fast-slow cascades can link
lightweight edge models with richer back-end models for video analytics, and sensor
processing, where the ``slow'' stage may
correspond to longer context, larger models, or different hardware.

%-------------------------------------------------------------------------------
\begin{acks}
We sincerely thank the reviewers and shepherd for their valuable feedback and
suggestions. This material is based upon work supported by the U.S. National
Science Foundation under Grant No. 2229876 (NSF AI Institute for Agent-based
Cyber Threat Intelligence and Operation, ACTION), as well as the Faculty of
Engineering Interdisciplinary Fund at the University of Hong Kong.
\end{acks}

\bibliographystyle{ACM-Reference-Format}
\bibliography{paper}

\clearpage
\appendix

\setcounter{topnumber}{5}
\setcounter{bottomnumber}{3}
\setcounter{totalnumber}{8}
\setcounter{dbltopnumber}{4}
\renewcommand{\topfraction}{0.95}
\renewcommand{\bottomfraction}{0.85}
\renewcommand{\textfraction}{0.05}
\renewcommand{\floatpagefraction}{0.85}
\renewcommand{\dbltopfraction}{0.95}
\renewcommand{\dblfloatpagefraction}{0.85}
\setlength{\textfloatsep}{4pt plus 1pt minus 2pt}
\setlength{\floatsep}{3pt plus 1pt minus 2pt}
\setlength{\intextsep}{4pt plus 1pt minus 2pt}
\setlength{\dbltextfloatsep}{4pt plus 1pt minus 2pt}
\setlength{\dblfloatsep}{3pt plus 1pt minus 2pt}
\setlength{\abovecaptionskip}{1pt}
\setlength{\belowcaptionskip}{0pt}
\captionsetup{font=footnotesize, skip=1pt}
\captionsetup[subfigure]{font=scriptsize, skip=0pt}
\setlist[itemize]{leftmargin=*, itemsep=0.05em, topsep=0.15em, parsep=0pt, partopsep=0pt}

\section{Implementation Details}\label{app:implementation-details}

\begin{algorithm}[H]
  \begin{minipage}{\columnwidth}
  \scriptsize
  \caption{Slope-based Per-Class Uncertainty Thresholds}
  \label{alg:per_class_thresholds}
  \begin{tabbing}
  \hspace{0.8em}\=\hspace{1.3em}\=\hspace{1.3em}\=\kill
  \textbf{function} GetPerClassSlope($D, U$)\\
  \> Initialize Results $R$\\
  \> \textbf{for each} Class Label $L$ in $D$ \textbf{do}\\
  \>\> Partition $D$ into Correct $C$ and Incorrect $I$ for $L$ \hfill $\triangleright$ Classify predictions\\
  \>\> Get Thresholds $T$ from $U$ quantiles for $L$\\
  \>\> Update $R$ with assigned portions for $C$ and $I$ \hfill $\triangleright$ Track portions\\
  \> \textbf{end for}\\
  \> Sort $R$ and calculate slope $\frac{\Delta I}{\Delta (C + I)}$ \hfill $\triangleright$ Threshold adjustment\\
  \> \textbf{return} $R$\\
  \textbf{end function}\\
  \textbf{function} GetPerClassThresholds($D, U$)\\
  \> $R = $ GetPerClassSlope($D, U$)\\
  \> Initialize $TD$ with max $U$ for each class \hfill $\triangleright$ Highest uncertainty\\
  \> Set $AllAssigned$ to 0\\
  \> \textbf{for} each increment \textbf{do}\\
  \>\> Add top item from $R$ to MaxSlopeQueue $MSQ$\\
  \>\> \textbf{if} $MSQ$ is empty \textbf{then} \textbf{break} \hfill $\triangleright$ All thresholds are processed\\
  \>\> Select max slope from $MSQ$ \hfill $\triangleright$ Pick steepest class\\
  \>\> Update $TD$ and $AllAssigned$\\
  \>\> Remove selected item from $R$\\
  \> \textbf{end for}\\
  \> \textbf{return} Thresholds from $TD$ \hfill $\triangleright$ Final thresholds\\
  \textbf{end function}
  \end{tabbing}
  \end{minipage}
\end{algorithm}

\begin{algorithm}[H]
  \begin{minipage}{\columnwidth}
  % \centering
  \scriptsize
  \caption{Universal Uncertainty Thresholds}
  \label{alg:universal_threshold}
  \begin{tabbing}
  \hspace{0.8em}\=\hspace{1.3em}\=\kill
  \textbf{function} GetUniversalThresholds($D$)\\
  \> Initialize Uncertainty List $U$\\
  \> \textbf{for each} $DataPoint$ in Dataset $D$ \textbf{do}\\
  \>\> Compute Uncertainty $u$ for $DataPoint$ \hfill $\triangleright$ $D$ is validation data\\
  \>\> Append $u$ to $U$\\
  \> \textbf{end for}\\
  \> Sort $U$ in ascending order\\
  \> Define Quantile Values $Q \gets \{q_1, q_2, \dots, q_n\}$ \hfill $\triangleright$ E.g., 0.01 to 0.99\\
  \> Initialize Threshold List $T$\\
  \> \textbf{for each} $q \in Q$ \textbf{do}\\
  \>\> Compute Quantile Index $qIndex \gets \lceil q \times \text{length of } U \rceil$\\
  \>\> Append $U[qIndex]$ to $T$\\
  \> \textbf{end for}\\
  \> \textbf{return} Threshold List $T$\\
  \textbf{end function}
  \end{tabbing}
  \end{minipage}
\end{algorithm}

Once wired to a network traffic tap, \sysname can be flexibly deployed in various network environments 
as long as it meets the necessary throughput and latency thresholds for 
real-time traffic analysis. This allows it to provide immediate 
insights across different network scales and use cases.
Our system integrates several key technologies to create a robust ML solution for network traffic analysis, leveraging the strengths of
nPrint~\cite{holland2021new} with PF\_RING~\cite{pfring2023} for efficient packet capture, EdgeServe~\cite{shaowang2023edgeserve} for low-latency
model serving, and a combination of Scikit-learn~\cite{scikit-learn}, Torch~\cite{paszke2019pytorch}, LightGBM~\cite{ke2017lightgbm}, and ONNX~\cite{onnxruntime} for
model training, evaluation, and deployment. This comprehensive setup enables
us to handle diverse computational needs. 

For packet capture and initial processing, we use nPrint in conjunction with
PF\_RING C APIs, enhancing our ability to capture network packets
quickly in an expressive manner. This forms the foundation of our data
preparation phase, ensuring that we have high-quality data for model training
and inference. PF\_RING\_PROMISC enables promiscuous mode on the network
interface, allowing it to capture all packets on the network segment it is
attached to, not just those addressed to it. Following the original design of
nPrint, we implement using PF\_RING APIs to get  1024 bits from IPv4, TCP,
and UDP headers in default nPrint. We further enabled dynamic feature subscriptions.
\rchange{Dynamic feature subscriptions let each stage request only the
packet-level features it needs, typically 200--400 informative features from the
1024-bit nPrint encoding.}

Our machine learning models range from simpler structures like Decision Trees,
(at least 15 samples per leaf for enhanced uncertainty characterization), Random Forest
to more complex configurations such as CNNs. For gradient boosting models, we
employ LightGBM (learning rate 0.03, number of leaves 128, feature fraction 0.9,
minimum data in leaf 3) and XGBoost (100 estimators), each
carefully tuned with specific parameters to optimize performance.  The architecture of our CNN model is
designed to process inputs effectively, featuring layers for convolution,
pooling, and fully connected operations, alongside dropout for regularization.
The model's structure is tailored to the sequential nature of network packet
data, with adjustments made based on the input data's dimensions. These models
are trained on prepared datasets, converted to ONNX, and then loaded into our
system, adhering to operator set version 12 and machine learning
operator set version 2 for compatibility. \change{For each application, we 
train five different models across packet depths ranging from 1 to 20, 
resulting in a total of 100 models with different complexities.}

For model serving, EdgeServe~\cite{shaowang2023edgeserve} plays an important role. As a low-latency streaming
system designed for decentralized prediction, it facilitates intra-edge message
routing and communication. This capability allows all nodes within the system to
produce and consume both data and predictions efficiently, supporting the
dynamic integration of multiple data streams for enhanced prediction accuracy.
EdgeServe's integration enables each model to process inputs independently,
beginning with the fastest model for the first packet of each flow. This model's
predictions are immediately used for decision-making unless deemed
unsatisfactory by our flow selection algorithm. In such cases, or when more
detailed analysis is required, the slower models take over, incorporating
additional features and subsequent packets. This approach, supported
by a queueing mechanism for managing asynchronous data arrival and a FIFO
strategy for queue clearance, ensures that our system not only meets the accuracy
requirements but also maintains efficiency and responsiveness across various
operational scenarios. For deployment using ONNX Runtime, we specify inter- and
intra-operation parallelism to be both 1.

\section{Evaluation Setting Details}
\label{app:evaluation-setting-details}
\label{app:baseline-details}

\paragraph{Testbed Configuration.} For our experiments, the datasets
are split into 50\% for training, 10\% for validation, and 40\% for
testing. We develop and test a range of models including Decision
Tree, Random Forest, LightGBM, XGBoost, and CNN, across scenarios
from handling the first 1 packet to the first 10 packets. These models
are later placed at different locations of \sysname for evaluations.

To accurately simulate network conditions and maintain the original
packet inter-arrival times, we modify the testing set by adjusting
time stamps and altering their five-tuple identifiers before merging
them into new pcap files. These modified flows range from 200 to 60,000
per second and are encapsulated into 60-second pcap files. These
files are then replayed (by using TCPReplay) to a dummy network
interface capable of supporting up to 10 Gbps, which is monitored
by \sysname instances through PF\_RING APIs.

The testbed is running on an AMD EPYC 7302P 16-core 32-thread
processor equipped with 64GB RAM, which is a single lab server.
\rchange{Thus, our current evaluation demonstrates single-server scale-up with
multiple consumers and heterogeneous workers.}
Each consumer is an instance subscribing to the incoming data stream.
To evaluate system performance, we intentionally
restrict the processing to a single consumer if not otherwise specified.

\begin{table}[H]
  \centering
  \setlength{\tabcolsep}{8pt}
  \renewcommand{\arraystretch}{1.08}
  \begin{minipage}{\columnwidth}
    \centering
    \small
    \captionsetup{width=0.86\columnwidth}
    \begin{tabular}{@{}lrr@{}}
        \toprule
        \textbf{\#Cons.} & \textbf{All CPU} & \textbf{Half CPU+Half GPU} \\
        \midrule
        \textbf{1}  & 4.319k & - \\
        \textbf{2}  & 8.805k & 8.620k \\
        \textbf{4}  & 16.692k & 16.484k \\
        \textbf{8}  & 31.114k & 30.651k \\
        \textbf{12} & 40.238k & 41.009k \\
        \textbf{16} & 48.505k & 47.040k \\
        \bottomrule
    \end{tabular}
    \caption{Maximum service rate (new classified flows per second) when we increase the number of consumers.}
    \label{tab:service_rates_multi}
  \end{minipage}%
  \hfill
\end{table}

\paragraph{Baseline Details.}

We do not include SmartNIC-based systems (\eg,
Homunculus~\cite{swamy2023homunculus} and LEO~\cite{jafri2024leo}) because they
require specialized hardware. \sysname's flow-collection optimization is
orthogonal to their designs.

\begin{itemize}
  \item \textbf{LEXNet}~\cite{fauvel2023lightweight}. \change{LEXNet is a
  recent work that optimizes model size and efficiency while preserving
  explainability for traffic classification. It uses the LERes block, a
  lower-cost residual block, and the LProto block, which learns a variable
  number of prototypes per class. LEXNet only requires packet size and
  direction. Although the original LEXNet waits for 20 packets to extract
  features, we use the same packet depth as the other baselines for a fair
  comparison.}
  \item \textbf{FastTraffic}~\cite{xu2023fasttraffic}. \change{FastTraffic is a
  lightweight deep learning method for traffic analysis on resource-constrained
  devices. It works at the IP packet level, truncates the informative portions
  of packets, and uses a text-like N-gram embedding with a three-layer MLP. We
  adapt this per-packet method to per-flow inference to avoid repeated
  computation on packets from the same flow.}
  \item \textbf{Best Effort}. An inference is applied once the flow
  features are fully extracted, allowing the model to immediately
  make a prediction. We implement this baseline with PF\_RING-supported nPrint
  extraction and an ONNX-optimized model. Both stages run in C++ within the same
  thread.
  \item \textbf{Queueing}. Extracted flow features are placed into an in-memory
  queue and processed in FIFO order. This implementation mirrors \sysname but
  removes the fast-slow architecture.
\end{itemize}

\section{Datasets}\label{app:data}
We present the detailed descriptions of the three datasets we have evaluated in Table~\ref{table:service_recognition_dataset},~\ref{table:device_identification_dataset}, and~\ref{table:vca_qoe}. 

\begin{table}[H]
  \centering
  \small
  \setlength{\tabcolsep}{8pt}
  \renewcommand{\arraystretch}{1.08}
  \captionsetup{width=0.82\columnwidth}
  \begin{tabular}{@{}llr@{}}
      \toprule
      \textbf{Service} & \textbf{Application} & \textbf{\# of Flows} \\
      \midrule
      \multirow{3}{*}{Video Conferencing~\cite{macmillan2021measuring}} & Zoom & 1312 \\
      & Google Meet & 1313 \\
      & MS Teams & 3886 \\
      \midrule
      \multirow{4}{*}{Video Streaming~\cite{bronzino2019inferring}} & Twitch & 1150 \\
      & Amazon & 1509 \\
      & YouTube & 2702 \\
      & Netflix & 4104 \\
      \midrule
      \multirow{3}{*}{Social Media} & Instagram & 873 \\
      & Twitter & 1260 \\
      & Facebook & 1477 \\
      \midrule
      Smart Home~\cite{liu2023amir} & Other & 3901 \\
      \midrule
      \textbf{Total} & - & \textbf{23487} \\
      \bottomrule
  \end{tabular}
  \caption{Service recognition dataset.}
  \label{table:service_recognition_dataset}
\end{table}

\begin{table}[H]
  \centering
  \small
  \setlength{\tabcolsep}{8pt}
  \renewcommand{\arraystretch}{1.08}
  \captionsetup{width=0.78\columnwidth}
  \begin{tabular}{@{}lr@{}}
      \toprule
      \textbf{Device} & \textbf{\# of Flows} \\
      \midrule
      Samsung Fridge & 3770 \\
      Nvidia Jetson Nano & 3770 \\
      Amazon Echo 3rd Gen & 3770 \\
      Google Home & 3770 \\
      Raspberry Pi 3 & 3770 \\
      SmartThings Dishwasher & 3770 \\
      Philips Lightbulb & 3770 \\
      GE Washer & 3770 \\
      GE Dryer & 3770 \\
      Other & 3770 \\
      LG Nexus 5 & 3057 \\
      Nest Camera & 2543 \\
      Bose SoundTouch 30 & 1875 \\
      TP-Link Router AC1750 & 1523 \\
      Samsung Galaxy J3 & 1215 \\
      TP-Link Smart Plug HS100 & 1124 \\
      iRobot Vacuum & 728 \\
      Apple MacBook Pro & 252 \\
      \midrule
      \textbf{Total} & \textbf{50017} \\
      \bottomrule
  \end{tabular}
  \caption{Device identification dataset.}
  \label{table:device_identification_dataset}
\end{table}

\begin{table}[H]
  \centering
  \small
  \setlength{\tabcolsep}{8pt}
  \renewcommand{\arraystretch}{1.08}
  \captionsetup{width=0.62\columnwidth}
  \begin{tabular}{@{}lr@{}}
      \toprule
      \textbf{VCA} & \textbf{Total \# of Seconds} \\
      \midrule
      Meet & 10189 \\
      Webex & 12331 \\
      Teams & 14408 \\
      \midrule
      \textbf{Total} & \textbf{36928} \\
      \bottomrule
  \end{tabular}
  \caption{Video conferencing application (VCA) QoE measurement dataset.}
  \label{table:vca_qoe}
\end{table}

\section{Extended Evaluations}\label{app:perf}
\begin{figure}[H]
  \centering
  \subfloat[Single-consumer service rate.\label{fig:iot_trp_comp}]{%
    \includegraphics[width=0.49\columnwidth]{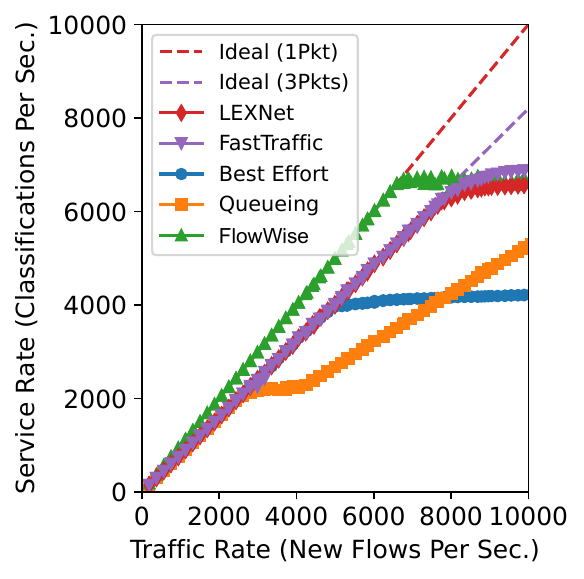}%
  }\hfill
  \subfloat[Average end-to-end latency.\label{fig:iot_se2e_t_comp}]{%
    \includegraphics[width=0.41\columnwidth]{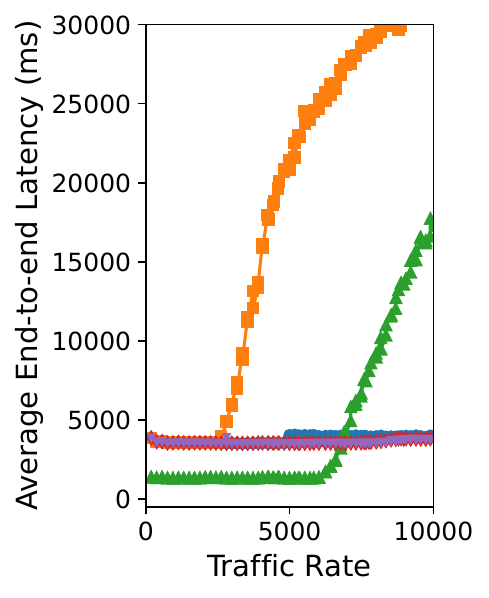}%
  }\\[0.2em]
  \subfloat[Miss rate (above 10Pkts).\label{fig:iot_long_lr_comp}]{%
    \includegraphics[width=0.45\columnwidth]{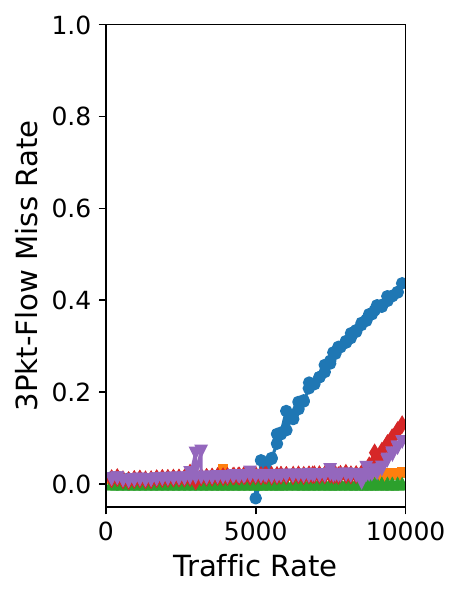}%
  }\hfill
  \subfloat[Max service rate vs. F1.\label{fig:iot_service_rate_f1}]{%
    \includegraphics[width=0.45\columnwidth]{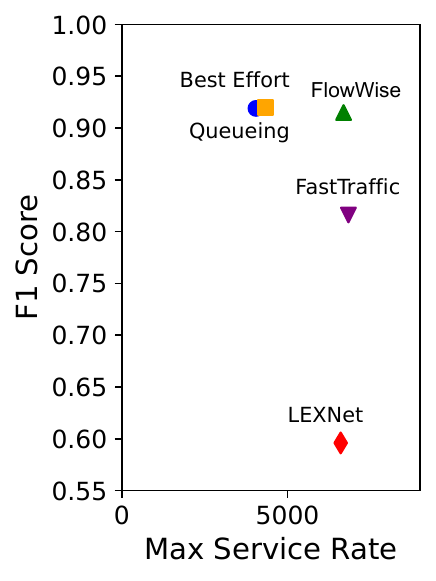}%
  }
  \caption{Performance of different approaches for the \textbf{device identification} task. We compare \sysname against 
  four baselines, all considering three-packet inputs. \change{The F1 score 
  is computed on the flows that are not missed.}}
  \label{fig:iot_comp}
\end{figure}

\paragraph{Service Rate, End-to-End Latency, and Miss Rate.}
For device identification, Figure~\ref{fig:iot_comp} shows that 
\sysname achieves better F1 score vs. service rate (6.69kfps) 
tradeoff than others. 
The four panels separate the sources of this tradeoff. The service-rate plot
compares the achieved classification rate against the offered traffic rate,
while the latency plot shows how long each completed decision spends in the
pipeline. The miss-rate plot captures the fraction of long flows that remain
unclassified after the three-packet budget, and the final scatter plot places
each method on the accuracy-throughput frontier. This breakdown is important
because a high raw service rate can hide delayed or missing decisions, which
are not useful for streaming traffic analysis.

Across these views, \sysname avoids the main failure modes of the baselines.
Queueing can preserve accuracy after the model receives enough features, but
the FIFO queue increases end-to-end latency as the offered traffic rate grows.
Best Effort avoids the extra queue, but its single late decision cannot exploit
easy flows that could have been classified earlier. FastTraffic and LEXNet are
lightweight, but their fixed inference structure does not adapt the amount of
work to each flow. \sysname instead assigns easy flows to faster stages and
keeps the slow stages for harder flows, so it maintains a stronger balance
between service rate, latency, miss rate, and F1 score.

\begin{figure}[H]
  \centering
  \subfloat[Fastest to fast.\label{fig:dev_id_f2f}]{%
    \includegraphics[width=0.48\columnwidth]{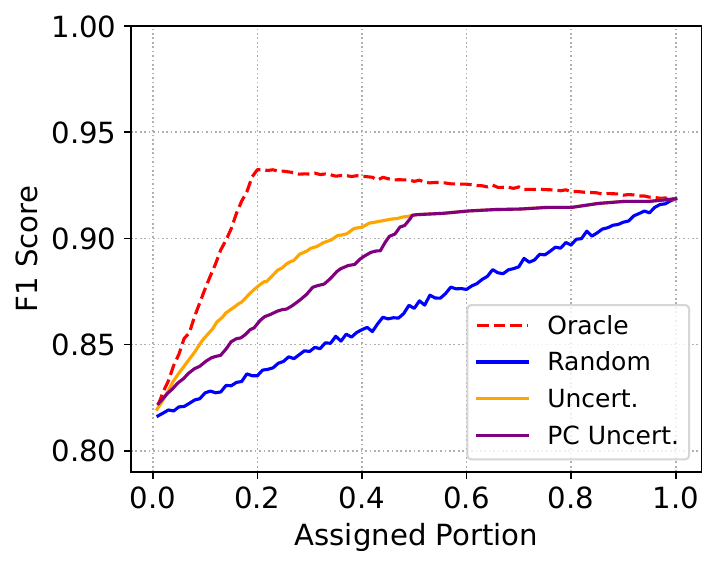}%
  }
  \hfill
  \subfloat[Fast to slow.\label{fig:dev_id_e2e}]{%
    \includegraphics[width=0.48\columnwidth]{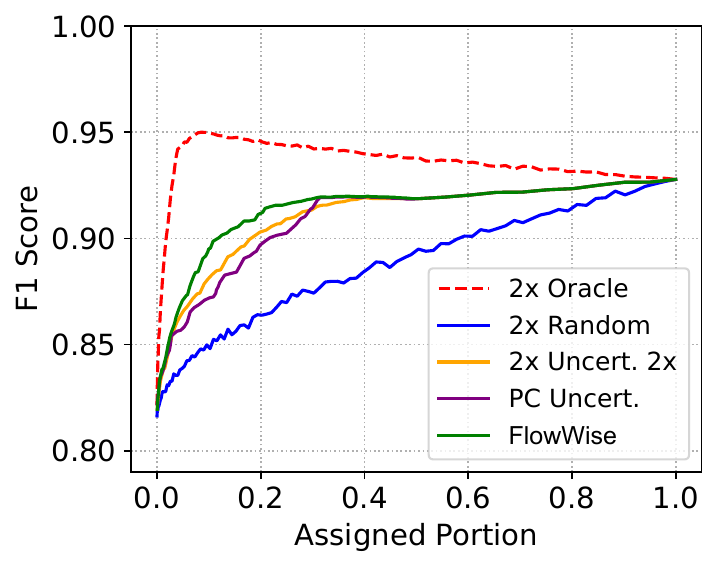}%
  }
  \caption{Assigned portion of requests vs. F1 score at each stage for device identification task.}
  \label{fig:dev_id_portion_f1}
\end{figure}

\begin{figure}[H]
  \centering
  \subfloat[Fastest to fast.\label{fig:vca_qoe_f2f}]{%
    \includegraphics[width=0.48\columnwidth]{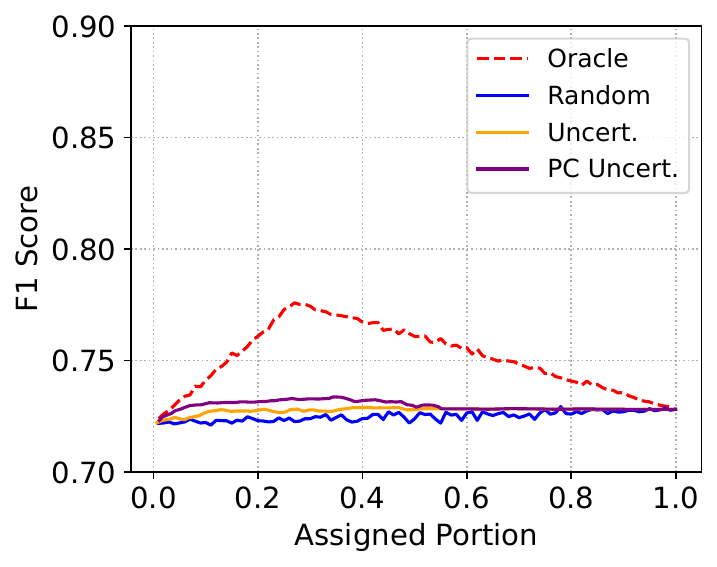}%
  }
  \hfill
  \subfloat[Fast to slow.\label{fig:vca_qoe_e2e}]{%
    \includegraphics[width=0.48\columnwidth]{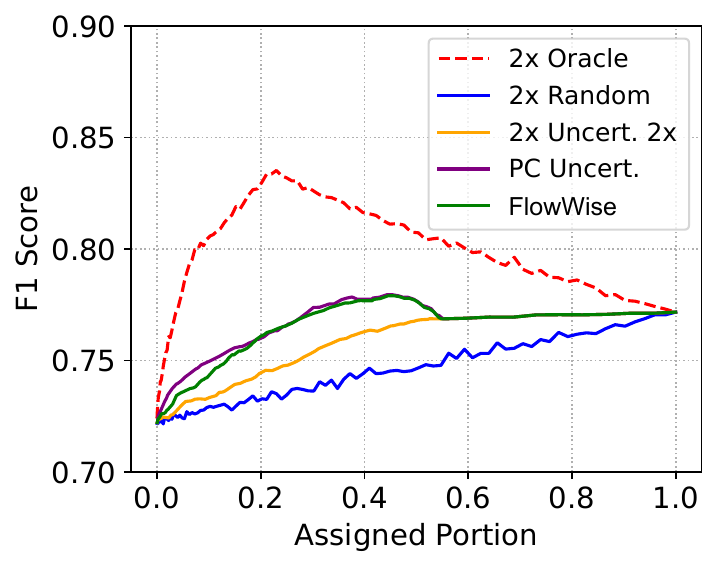}%
  }
  \caption{Assigned portion of requests vs. F1 score at each stage for QoE measurement task.}
  \label{fig:vca_qoe_portion_f1}
\end{figure}

\paragraph{Assigned Portion vs. F1 Score.} Figures~\ref{fig:dev_id_portion_f1} and~\ref{fig:vca_qoe_portion_f1} show the effectiveness of assignment algorithms for device identification and QoE measurement tasks.
The x-axis controls how much traffic is assigned to an earlier stage, and the
y-axis reports the resulting F1 score after that assignment. The oracle curve
gives an upper-bound reference by assigning flows with hindsight. Random
assignment gives a lower-bound reference because it does not distinguish easy
flows from ambiguous flows. The uncertainty-based policies sit between these
two references and use validation-set uncertainty to approximate which flows
can leave the pipeline early.

The device-identification task benefits more from uncertainty-guided assignment
because the device classes have more separable early-packet signatures. In this
setting, the Per-Class Uncertainty policy can preserve high F1 while increasing
the assigned portion. The QoE task is harder because the labels are tied to
application quality rather than only endpoint identity, so the curves are
closer together and the gains from early assignment are smaller. These results
support the design choice of keeping assignment policies configurable across
tasks instead of fixing one thresholding rule for all deployments.
They also show why the assignment objective must consider the whole pipeline.
Assigning too many flows early can improve throughput but can also remove the
examples for which later stages provide the largest accuracy gain. Conversely,
assigning too few flows early leaves capacity unused at the faster stages. The
uncertainty policies provide a practical middle ground by turning calibration
behavior into per-task assignment rules.

\paragraph{Multi-Consumer Throughput across Hardware Types.} We also evaluated the service rate using heterogeneous hardware, specifically a
combination of half CPU and half GPU consumers. The performance achieved was
similar to that of all CPU consumers, with slightly lower results due to the
overhead of moving features from RAM to GPU VRAM. Additionally, \sysname
currently runs one prediction at a time, which does not fully utilize the
multi-threading capabilities of GPUs. Future work could explore more advanced
batching techniques to address this limitation.
Table~\ref{tab:service_rates_multi} also shows that the consumer abstraction
scales beyond a single worker. Increasing the number of consumers raises the
maximum service rate from 4.319k classified flows per second with one consumer
to 48.505k classified flows per second with sixteen CPU consumers. The scaling
is not perfectly linear because consumers still share packet capture, feature
extraction, memory bandwidth, and stream-routing costs. Even so, the results
show that \sysname can use additional consumers to raise throughput without
changing the model-placement policy.
Operators can add consumers as traffic volume changes while keeping the same
fast-slow model sequence and flow-selection thresholds.

%%%%%%%%%%%%%%%%%%%%%%%%%%%%%%%%%%%%%%%%%%%%%%%%%%%%%%%%%%%%%%%%%%%%%%%%%%%%%%%%
\end{document}